\pdfoutput=1
\documentclass[bib]{statapress}

\usepackage[crop,newcenter,frame]{pagedims}

\usepackage{bm} 
\usepackage{bbm}
\usepackage{amsmath}
\usepackage{amssymb}
\usepackage{amsfonts}
\usepackage{blkarray,bigstrut} 
\usepackage{subcaption}
\usepackage{float}
\usepackage{enumitem}
\usepackage{multirow}
\usepackage{tabularx}
\usepackage{xspace}
\usepackage{xcolor}
\usepackage{booktabs}
\usepackage{setspace}
\usepackage[normalem]{ulem} 

\usepackage{sj}

\usepackage{epsfig}

\usepackage{stata}

\usepackage{shadow}

\sjsetissue{$vv$}{$ii$}{$mm$}{$yyyy$}

\begin{document}


\graphicspath{{fig/}}

\makeatletter
\def\input@path{{tab/}{mc/}}
\makeatother


\newcommand{\sjversion}[2]{#2}

\newcommand{\norm}[1]{\left\lVert#1\right\rVert}
\newcommand{\normline}[1]{\lVert#1\rVert}

\newcommand{\myoption}[2]{\texttt{#1(}#2\texttt{)}}
\newcommand{\ttt}[1]{\texttt{#1}}
\newcommand{\lassopack}{\textbf{lassopack}\xspace}
\newcommand{\pdslasso}{\textbf{pdslasso}\xspace}
\newcommand{\lassotwo}{\stcmd{lasso2}\xspace}
\newcommand{\cvlasso}{\stcmd{cvlasso}\xspace}
\newcommand{\rlasso}{\stcmd{rlasso}\xspace}
\newcommand{\lasso}{lasso\xspace}
\newcommand{\obs}{n}
\newcommand{\pload}{\psi}
\newcommand{\Pload}{\psi}

\newcommand{\real}{{\it real}}
\newcommand{\stint}{{\it integer}}
\newcommand{\stvarname}{{\it varname}}
\newcommand{\var}{{\it variable}}
\newcommand{\method}{{\it method}}
\newcommand{\stnumlist}{{\it numlist}}
\newcommand{\stmatrix}{{\it matrix}}

\newcommand{\achim}[1]{\textcolor{red}{$<<$AA: #1$>>$}}
\newcommand{\ms}[1]{\textcolor{blue}{$<<$MS: #1$>>$}}
\newcommand{\chris}[1]{\textcolor{violet}{$<<$CH: #1$>>$}}


\inserttype[st0001]{article}
\author{Ahrens, Hansen \& Schaffer}{%
  Achim Ahrens\\The Economic and Social Research Institute\\Dublin, Ireland\\achim.ahrens@esri.ie
  \and
  Christian B. Hansen\\  University of Chicago\\christian.hansen@chicagobooth.edu
  \and
 Mark E. Schaffer\\ Heriot-Watt University\\Edinburgh, United Kingdom \\ m.e.schaffer@hw.ac.uk
}
\title[lassopack]{lassopack: Model selection and prediction with regularized regression in Stata}

\maketitle
\sjversion{}{\thispagestyle{empty}}

\begin{abstract}
This article introduces \lassopack, a suite of programs for regularized regression in Stata. \lassopack implements  lasso, square-root lasso, elastic net, ridge regression, adaptive lasso and post-estimation OLS. The methods are suitable for the high-dimensional setting where the number of predictors $p$ may be large and possibly greater than the number of observations, $\obs$. 
We offer three different approaches for selecting the penalization (`tuning') parameters: information criteria (implemented in \lassotwo), $K$-fold cross-validation and $h$-step ahead rolling cross-validation for cross-section, panel and time-series data (\cvlasso), and theory-driven (`rigorous') penalization for the lasso and square-root lasso for cross-section and panel data (\rlasso). 
We discuss the theoretical framework and practical considerations for each approach. We also present Monte Carlo results to compare the performance of the penalization approaches.

\keywords{\sjversion{\inserttag, }{}lasso2, cvlasso, rlasso, lasso, elastic net, square-root lasso, cross-validation}
\end{abstract}


\section{Introduction}
Machine learning is attracting increasing attention across a wide range of scientific disciplines. Recent surveys explore how machine learning methods can be utilized in economics and applied econometrics \citep{Varian2014,Mullainathan2017,Athey2018,Kleinberg2018}. 
At the same time, Stata offers to date only a limited set of machine learning tools. \lassopack is an attempt to fill this gap by providing easy-to-use and flexible methods for regularized regression in Stata.\footnote{This article refers to version 1.2 of \lassopack released on the 15th of January, 2019. For additional information and data files, see  \url{https://statalasso.github.io/}.}

While regularized linear regression is only one of many  methods in the toolbox of machine learning, it has some properties that make it attractive for empirical research. To begin with, it is a straightforward extension of linear regression. Just like ordinary least squares (OLS), regularized linear regression minimizes the sum of squared deviations between observed and model predicted values, but imposes a regularization penalty aimed at limiting model complexity. The most popular regularized regression method is the lasso---which this package is named after---introduced by \citet{Frank1993} and \citet{Tibshirani1996}, which penalizes the absolute size of coefficient estimates.

The primary purpose of regularized regression, like supervised machine learning methods more generally, is prediction.
Regularized regression typically does not produce estimates that can be interpreted as causal and statistical inference on these coefficients is complicated.\footnote{This is an active area of research, see for example \citet{Buhlmann2013,Meinshausen2009b,Weilenmann2017,Wasserman2009,Lockhart2014}.\label{fn:inference_hdm}} While regularized regression may select the true model as the sample size increases, this is generally only the case under strong assumptions.
However, regularized regression can aid causal inference without relying on the strong assumptions required for perfect model selection. The post-double-selection methodology of \citet{Belloni2014b} and the post-regularization approach of \citet{Chernozhukov2015} can be used to select appropriate control variables from a large set of putative confounding factors and, thereby, improve robustness of estimation of the parameters of interest. Likewise, the first stage of two-step least-squares is a prediction problem and lasso or ridge can be applied to obtain optimal instruments \citep{Belloni2012,Carrasco2012,Hansen2014}. These methods are implemented in our sister package \pdslasso \citep{pdslasso2018}, which builds on the algorithms developed in \lassopack.

The strength of regularized regression as a prediction technique stems from the \emph{bias-variance trade-off}. The prediction error can be decomposed into the unknown error variance reflecting the overall noise level (which is irreducible), the squared estimation bias and the variance of the predictor. The variance of the estimated predictor is increasing in the model complexity, whereas the bias tends to decrease with model complexity. By reducing model complexity and inducing a shrinkage bias, regularized regression methods tend to outperform OLS in terms of \emph{out-of-sample} prediction performance. In doing so, regularized regression addresses the common problem of overfitting: high in-sample fit (high $R^2$), but poor prediction performance on unseen data.

Another advantage is that the regularization methods of \lassopack---with the exception of ridge regression---are able to produce sparse solutions and, thus, can serve as model selection techniques. 
Especially when faced with a large number of putative predictors, model selection is challenging.
Iterative testing procedures, such as the \emph{general-to-specific approach}, typically induce pre-testing biases and hypothesis tests often lead to many false positives.
At the same time, high-dimensional problems where the number of predictors is large relative to the sample size are a common phenomenon, especially when the true model is treated as unknown. 
Regularized regression is well-suited for high-dimensional data. The $\ell_1$-penalization can set some coefficients to exactly zero, thereby excluding predictors from the model. 
The \emph{bet on sparsity} principle allows for identification even when the number of predictors exceeds the sample size under the assumption that the true model is sparse or can be approximated by a sparse parameter vector.\footnote{\citet[][p.~611]{Hastie2009} summarize the \emph{bet on sparsity} principle as follows: `Use a procedure that does well in sparse problems, since no procedure does well in dense problems.'}

Regularized regression methods rely on tuning parameters that control the degree and type of penalization.  \lassopack offers three approaches to select these tuning parameters. The classical approach is to select tuning parameters using cross-validation in order to optimize \emph{out-of-sample} prediction performance. Cross-validation methods are universally applicable and generally perform well for prediction tasks, but are computationally expensive.
A second approach relies on information criteria such as the \emph{Akaike information criterion} \citep{Zou2007,Zhang2010}. Information criteria are easy to calculate and have attractive theoretical properties, but are less robust to violations of the independence and homoskedasticity assumptions \citep{Arlot2010}. 
Rigorous penalization for the lasso and square-root lasso provides a third option. The approach is valid in the presence of heteroskedastic, non-Gaussian and cluster-dependent errors \citep{Belloni2012,Belloni2014c,Kozbur2015}. The rigorous approach places a high priority on controlling overfitting, thus often producing parsimonious models. This strong focus on containing overfitting is of practical and theoretical benefit for selecting control variables or instruments in a structural model, but also implies that the approach may be outperformed by cross-validation techniques for pure prediction tasks.
Which approach is most appropriate depends on the type of data at hand and the purpose of the analysis. To provide guidance for applied reseachers, we discuss the theoretical foundation of all three approaches, and present Monte Carlo results that assess their relative performance.

The article proceeds as follows. In Section~\ref{sec:estimators}, we present the estimation methods implemented in \lassopack. Section~\ref{sec:informationcriteria}-\ref{sec:rigorous} discuss the aforementioned approaches for selecting the tuning parameters: information criteria in Section~\ref{sec:informationcriteria}, cross-validation in Section~\ref{sec:crossvalidation} and rigorous penalization in Section~\ref{sec:rigorous}. The three commands, which correspond to the three penalization approaches, are presented in Section~\ref{sec:commands}, followed by demonstrations in Section~\ref{sec:demo}. Section~\ref{sec:mc} presents Monte Carlo results. Further technical notes are in Section~\ref{sec:technical}.


\paragraph{Notation.} We briefly clarify the notation used in this article. Suppose $\bm{a}$ is a vector of dimension $m$ with typical element $a_j$ for $j=1,\ldots,m$. The $\ell_1$-norm is defined as $\norm{\bm{a}}_1 = \sum_{j=1}^m |a_j|$, and the $\ell_2$-norm is $\norm{\bm{a}}_2 = \sqrt{\sum_{j=1}^m |a_j|^2}$. The `$\ell_0$-norm' of $\bm{a}$ is denoted by $\norm{a}_0$ and is equal to the number of non-zero elements in $\bm{a}$. $\mathbbm{1}\{.\}$ denotes the indicator function. We use the notation $b\vee c$ to denote the maximum value of $b$ and~$c$, i.e., $\max(b,c)$.




\section{Regularized regression}\label{sec:estimators}
This section introduces the regularized regression methods implemented in \lassopack. 
We consider the high-dimensional linear model \[y_i=\bm{x}_i'\bm{\beta}+\varepsilon_i,\qquad i=1,\ldots, n,\] where the number of predictors, $p$, may be large and even exceed the sample size, $n$.     
The regularization methods introduced in this section can accommodate large-$p$ models under the assumption of sparsity: out of the $p$ predictors only a subset of $s\ll n$ are included in the true model where $s$ is the sparsity index \[s := \sum_{j=1}^p \mathbbm{1}\{\beta_j\neq 0\}=\norm{\bm\beta}_0.\] 
We refer to this assumption as \emph{exact sparsity}. It is more restrictive than required, but we use it here for illustrative purposes. We will later relax the assumption to allow for non-zero, but `small', $\beta_j$ coefficients. We also define the active set $\Omega=\{j \in \{1,\ldots,p \}: \beta_j\neq 0\}$, which is the set of non-zero coefficients. In general, $p$, $s$, $\Omega$ and $\bm{\beta}$ may depend on $n$ but we suppress the $n$-subscript for notational convenience. 

We adopt the following convention throughout the article: unless otherwise noted, all variables have been mean-centered such that $\sum_i y_{i}=0$ and $\sum_ix_{ij}=0$, and all variables are measured in their natural units, i.e., they have \emph{not} been pre-standardized to have unit variance.
By assuming the data have already been mean-centered we simplify the notation and exposition. Leaving the data in natural units, on the other hand, allows us to discuss standardization in the context of penalization. 

Penalized regression methods rely on tuning parameters that control the degree and type of penalization.
The estimation methods implemented in \lassopack, which we will introduce in the following sub-section, use two tuning parameters: $\lambda$ controls the general degree of penalization and $\alpha$ determines the relative contribution of $\ell_1$ vs. $\ell_2$ penalization.
The three approaches offered by \lassopack for selecting $\lambda$ and $\alpha$ are introduced in \ref{sec:choice_of_lambda}.


\subsection{The estimators}\label{sec:estimators_subsection}

\subsubsection{Lasso}\label{sec:lasso}
 The lasso takes a special position, as it provides the basis for the rigorous penalization approach (see Section~\ref{sec:rigorous}) and has inspired other methods such as elastic net and square-root lasso, which are introduced later in this section. The lasso minimizes the  mean squared error subject to a penalty on the absolute size of coefficient estimates:
\begin{equation}
\bm{\hat\beta}_{\textrm{lasso}}(\lambda)=
\arg\min \frac{1}{n} \sum_{i=1}^n\left( y_i - \bm{x}_i'\bm{\beta} \right)^2 
    + \frac{\lambda}{n} \sum_{j=1}^p\psi_j|\beta_j|. 
    \label{eq:lasso}
\end{equation}
The tuning parameter $\lambda$ controls the overall penalty level and $\psi_j$ are predictor-specific penalty loadings.

\citet{Tibshirani1996} motivates the lasso with two major advantages over OLS.  First, due to the nature of the $\ell_1$-penalty, the lasso sets some of the coefficient estimates exactly to zero and, in doing so, removes some predictors from the model. Thus, the lasso serves as a model selection technique and facilitates model interpretation. Secondly, lasso can outperform least squares in terms of prediction accuracy due to the bias-variance trade-off. 

The lasso coefficient path, which constitutes the trajectory of coefficient estimates as a function of $\lambda$, is piecewise linear with changes in slope where variables enter or leave the active set. The change points are referred to as \emph{knots}. $\lambda=0$ yields the OLS solution and $\lambda\rightarrow\infty$ yields an empty model, where all coefficients are zero.

The lasso, unlike OLS, is not invariant to linear transformations, which is why scaling matters. If the predictors are not of equal variance, the most common approach is to pre-standardize the data such that $\frac{1}{\obs}\sum_i x^2_{ij}=1$ and set $\psi_j=1$ for $j=1,\ldots,p$. Alternatively, we can set the penalty loadings to $\hat\psi_j=(\frac{1}{\obs}\sum_ix_{ij}^2)^{-1/2}.$
The two methods yield identical results in theory.


\subsubsection{Ridge regression}
Ridge regression \citep{Tikhonov1963,Hoerl1970} replaces the $\ell_1$-penalty of the lasso with a $\ell_2$-penalty, thus minimizing
\begin{equation}
 \frac{1}{n} \sum_{i=1}^n\left( y_i - \bm{x}_i'\bm{\beta} \right)^2 
    +    \frac{\lambda}{n} \sum_{j=1}^p\psi_j^2 \beta_j^2.
\end{equation}
The interpretation and choice of the penalty loadings $\psi_j$ is the same as above. As in the case of the lasso, we need to account for uneven variance, either through pre-estimation standardization or by appropriately choosing the penalty loadings $\psi_j$.

In contrast to estimators relying on $\ell_1$-penalization, the ridge does not perform variable selection. At the same time, it also does not rely on the assumption of sparsity. This makes the ridge attractive in the presence of dense signals, i.e., when the assumption of sparsity does not seem plausible. 
Dense high-dimensional problems are more challenging than sparse problems: for example, \citet{Dicker2016} shows that, if $p/n\rightarrow\infty$, it is not possible to outperform a trivial estimator that only includes the constant. 
If $p,n\rightarrow$ jointly, but $p/n$ converges to a finite constant, the ridge has desirable properties in dense models and tends to perform better than sparsity-based methods \citep{Hsu2014,Dicker2016,Dobriban2018}.

Ridge regression is closely linked to principal component regression. Both methods are popular in the context of multicollinearity due to their low variance relative to OLS. Principal components regression applies OLS to a subset of components derived from principal component analysis; thereby discarding a specified number of components with low variance. The rationale for removing low-variance components is that the predictive power of each component tends to increase with the variance. 
The ridge can be interpreted as projecting the response against principal components while imposing a higher penalty on components exhibiting low variance. Hence, the ridge follows a similar principle; but, rather than discarding low-variance components, it applies a more severe shrinkage \citep{Hastie2009}.



A comparison of lasso and ridge regression provides further insights into the nature of $\ell_1$ and $\ell_2$ penalization. For this purpose, it is helpful to write lasso and ridge in constrained form as
\begin{align*}
 \bm{\hat\beta}_{\textrm{lasso}} &=  \arg\min \frac{1}{n} \sum_{i=1}^p\left( y_i - \bm{x}_i'\bm{\beta} \right)^2 
    \quad\textrm{subject to}\quad  \sum_{j=1}^n\psi_j |\beta_j| \leq \tau, \\
\bm{\hat\beta}_{\textrm{ridge}} &=   \arg\min   \frac{1}{n} \sum_{i=1}^n\left( y_i - \bm{x}_i'\bm{\beta} \right)^2 
    \quad\textrm{subject to}\quad  \sum_{j=1}^p\psi_j^2 \beta_j^2 \leq \tau 
\end{align*}
and to examine the shapes of the constraint sets. The above optimization problems use the tuning parameter $\tau$ instead of $\lambda$. Note that there exists a data-dependent relationship between $\lambda$ and $\tau$.

\begin{figure}
    \centering
    \begin{minipage}{.4\linewidth}
    \sjversion{\includegraphics[width=\linewidth]{lasso-v-ridge_mono.pdf}}
    {\includegraphics[width=\linewidth]{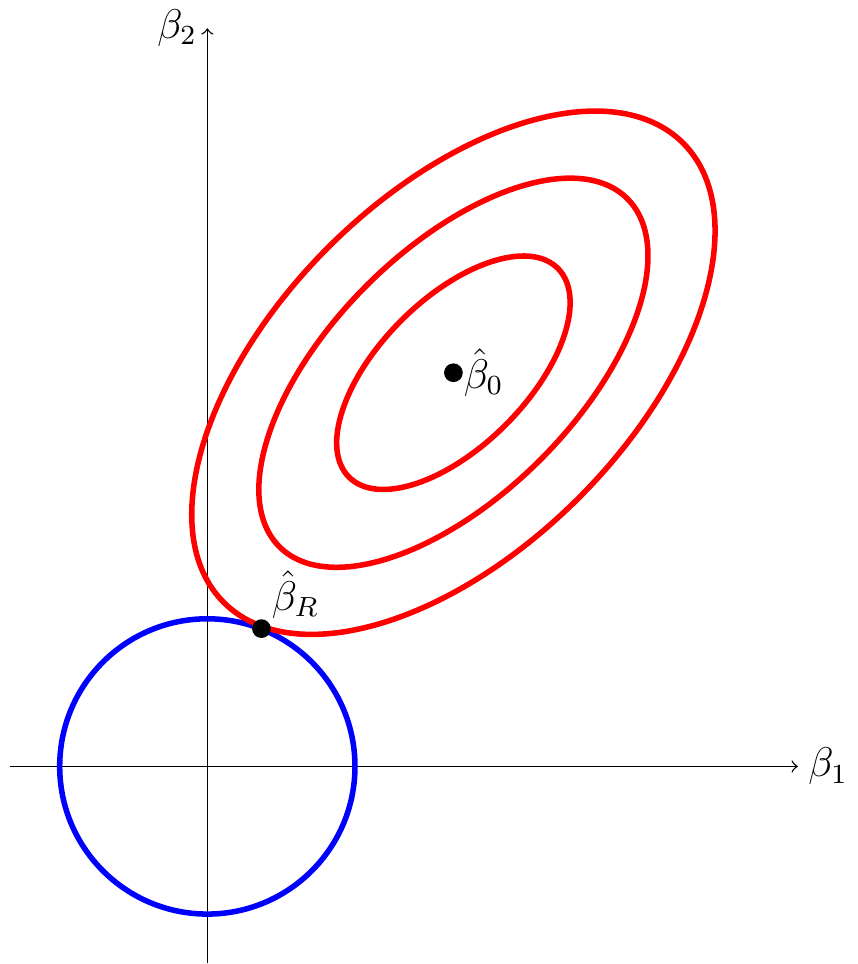}}
    \subcaption{Ridge}
    \end{minipage}%
    \begin{minipage}{.4\linewidth}
    \sjversion{\includegraphics[width=\linewidth]{lasso-v-ridge2_mono.pdf}}{\includegraphics[width=\linewidth]{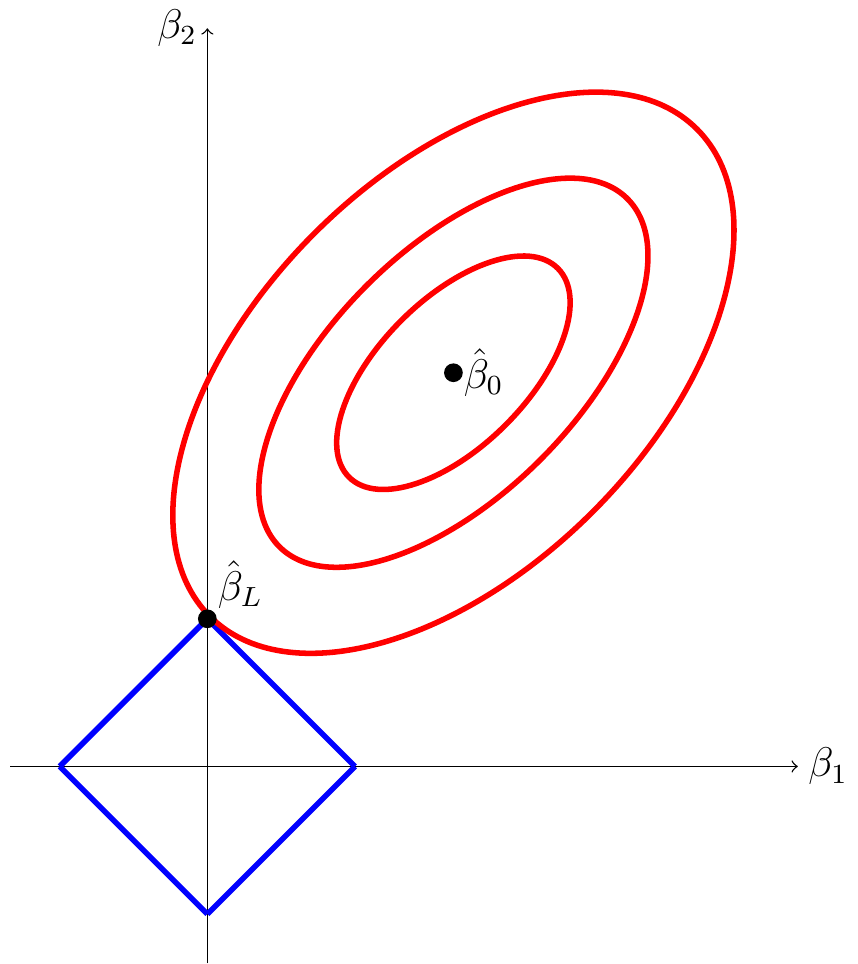}}
    \subcaption{Lasso}
    \end{minipage}
    \caption{Behaviour of $\ell_1$ and $\ell_2$-penalty in comparison. \sjversion{Gray}{Red} lines represent RSS contour lines and the \sjversion{black}{blue} lines represent the lasso and ridge constraint, respectively. $\hat\beta_0$ denotes the OLS estimate. $\hat\beta_L$ and $\hat\beta_R$ are the lasso and ridge estimate. The illustration is based on Tibshirani, 1996, Fig. 2.
    }
    \label{fig:lasso_v_ridge}
\end{figure}

Figure~\ref{fig:lasso_v_ridge} illustrates the geometry underpinning lasso and ridge regression for the case of $p=2$ and $\psi_1=\psi_2=1$ (i.e., unity penalty loadings). The \sjversion{grey}{red} elliptical lines represent residual sum of squares contours and the \sjversion{black}{blue} lines indicate the lasso and ridge constraints. The lasso constraint set, given by $|\beta_1| + |\beta_2| \leq \tau$, is diamond-shaped with vertices along the axes from which it immediately follows that the lasso solution may set coefficients exactly to 0.  In contrast, the ridge constraint set, $\beta_1^2 + \beta_2^2  \leq \tau$, is circular and will thus (effectively) never produce a solution with any coefficient set to 0. Finally, $\hat\beta_0$ in the figure denotes the solution without penalization, which corresponds to OLS. The lasso solution at the corner of the diamond implies that, in this example, one of the coefficients is set to zero, whereas ridge and OLS produce non-zero estimates for both coefficients.

While there exists no closed form solution for the lasso, the ridge solution can be expressed as \[ \bm{\hat\beta}_{\textrm{ridge}} = (\bm{X}'\bm{X}+\lambda\bm{\Psi}'\bm{\Psi})^{-1}\bm{X}'\bm{y}. \]
Here $\bm{X}$ is the $\obs \times p$ matrix of predictors with typical element $x_{ij}$, $\bm{y}$ is the response vector and $\bm{\Psi}=\textrm{diag}(\psi_1,\ldots,\psi_p)$ is the diagonal matrix of penalty loadings.  The ridge regularizes the regressor matrix by adding positive constants to the diagonal of $\bm{X}'\bm{X}$. The ridge solution is thus well-defined generally as long as all the $\psi_j$ and $\lambda$ are sufficiently large even if $\bm{X}'\bm{X}$ is rank-deficient.

\subsubsection{Elastic net}
The elastic net of \citet{ZouHastie2005} combines some of the strengths of lasso and ridge regression. It applies a mix of $\ell_1$ (lasso-type) and $\ell_2$ (ridge-type) penalization:
\begin{equation}
 \bm{\hat\beta}_\textrm{elastic}=\arg\min\frac{1}{n} \sum_{i=1}^n\left( y_i - \bm{x}_i'\bm{\beta} \right)^2 
    + \frac{\lambda}{n} \left[ \alpha \sum_{j=1}^p\psi_j|\beta_j|
    + (1-\alpha)\sum_{j=1}^p\psi_j^2\beta_j^2 \right]
\end{equation}
The additional parameter $\alpha$ determines the relative to contribution of $\ell_1$ vs. $\ell_2$ penalization. In the presence of groups of correlated regressors, the lasso typically selects only one variable from each group, whereas the ridge tends to produce similar coefficient estimates for groups of correlated variables. On the other hand, the ridge does not yield sparse solutions impeding model interpretation. The elastic net is able to produce sparse solutions for some $\alpha$ greater than zero, and retains or drops correlated variables jointly.

\subsubsection{Adaptive lasso}
The \emph{irrepresentable condition (IRC)} is shown to be sufficient and (almost) necessary for the lasso to be model selection consistent \citep{Zhao2006,Meinshausen2006}. However, the IRC imposes strict constraints on the degree of correlation between predictors in the true model and predictors outside of the model. Motivated by this non-trivial condition for the lasso to be variable selection consistent, \citet{Zou2006} proposed the adaptive lasso. The adaptive lasso uses penalty loadings of $\psi_j=1/|\hat\beta_{0,j}|^\theta$ where $\hat\beta_{0,j}$ is an initial estimator.  The adaptive lasso is variable-selection consistent for fixed $p$ under weaker assumptions than the standard lasso.  If $p<n$, OLS can be used as the initial estimator.  \citet{Huang2008} prove variable selection consistency for large $p$ and suggest using univariate OLS if $p>n$. The idea of adaptive penalty loadings can also be applied to elastic net and ridge regression \citep{Zou2009}.

\subsubsection{Square-root lasso}
The square-root lasso,
\begin{equation}
\bm{\hat\beta}_{\sqrt{\textrm{lasso}}} =
\arg\min \sqrt{\frac{1}{n} \sum_{i=1}^n\left( y_i - \bm{x}_i'\bm{\beta} \right)^2} 
    + \frac{\lambda}{n} \sum_{j=1}^p\psi_j|\beta_j|,
\end{equation}
is a modification of the lasso that minimizes the root mean squared error, while also imposing an $\ell_1$-penalty. The main advantage of the square-root lasso over the standard lasso becomes apparent if theoretically grounded, data-driven penalization is used. Specifically, the score vector, and thus the optimal penalty level, is independent of the unknown error variance under homoskedasticity as shown by \citet{Belloni2011a},
resulting in a simpler procedure for choosing $\lambda$
(see Section~\ref{sec:rigorous}).

\subsubsection{Post-estimation OLS}
Penalized regression methods induce an attenuation bias that can be alleviated by post-estimation OLS, which applies OLS to the variables selected by the first-stage variable selection method, i.e., 
\begin{equation}
\bm{\hat\beta}_{\textrm{post}} =
\arg\min
 \frac{1}{n} \sum_{i=1}^n\left( y_i - \bm{x}_i'\bm{\beta} \right)^2 
    \qquad \textrm{subject to}\qquad \beta_j = 0~~ \textrm{if}~~ \tilde\beta_j=0,
\end{equation}
where $\tilde\beta_j$ is a sparse first-step estimator such as the lasso. Thus, post-estimation OLS treats the first-step estimator as a genuine model selection technique. For the case of the lasso, \citet{Belloni2013} have shown that the post-estimation OLS, also referred to as post-lasso, performs at least as well as the lasso under mild additional assumptions if theory-driven penalization is employed. Similar results hold for the square-root lasso \citep{Belloni2011a,Belloni2014c}.



\subsection{Choice of the tuning parameters}\label{sec:choice_of_lambda}
Since coefficient estimates and the set of selected variables depend on $\lambda$ and $\alpha$, a central question is how to choose these tuning parameters. 
Which method is most appropriate depends on the objectives and setting: 
in particular, the aim of the analysis (prediction or model identification), computational constraints, and if and how the i.i.d.\ assumption is violated.
\lassopack offers three approaches for selecting the penalty level of $\lambda$ and~$\alpha$:
\begin{enumerate} 
\item \emph{Information criteria:} The value of $\lambda$ can be selected using information criteria. 
    \lassotwo implements model selection using four information criteria.
    We discuss this approach in Section~\ref{sec:informationcriteria}.
    \item \emph{Cross-validation:} The aim of cross-validation is to optimize the out-of-sample prediction performance.
    Cross-validation is implemented in \cvlasso, which allows for cross-validation across both $\lambda$ and the elastic net parameter $\alpha$. See Section~\ref{sec:crossvalidation}.
    \item \emph{Theory-driven (`rigorous'):} 
    Theoretically justified and feasible penalty levels and loadings are available for the lasso and square-root lasso via \rlasso.
    The penalization is chosen to dominate the noise of the data-generating process (represented by the score vector), which allows derivation of theoretical results with regard to consistent prediction and parameter estimation. See Section~\ref{sec:rigorous}.
\end{enumerate}  


\section{Tuning parameter selection using information criteria}\label{sec:informationcriteria}
Information criteria are closely related to regularization methods. The classical \emph{Akaike's information criterion} \citep[][AIC]{Akaike1974} is defined as $-2 \times \textnormal{log-likelihood} + 2 p$. Thus, the AIC can be interpreted as penalized likelihood which imposes a penalty on the number of predictors included in the model. This form of penalization, referred to as $\ell_0$-penalty, has, however, an important practical disadvantage. In order to find the model with the lowest AIC, we need to estimate all different model specifications. 
In practice, it is often not feasible to consider the full model space. For example, with only 20 predictors, there are more than 1~million different models. 

The advantage of regularized regression is that it provides a data-driven method for reducing model selection to a one-dimensional problem (or two-dimensional problem in the case of the elastic net) where we need to select $\lambda$ (and $\alpha)$. Theoretical properties of information criteria are well-understood and they are easy to compute once coefficient estimates are obtained. Thus, it seems natural to utilize the strengths of information criteria as model selection procedures to select the penalization level.  

Information criteria can be categorized based on two central properties: \emph{loss efficiency} and \emph{model selection consistency}.  A model selection procedure is referred to as loss efficient if it yields the smallest averaged squared error attainable by all candidate models. Model selection consistency requires that the true model is selected with probability approaching 1 as $n\rightarrow\infty$. Accordingly, which information information criteria is appropriate in a given setting also depends on whether the aim of analysis is prediction or identification of the true model. 

We first consider the most popular information criteria, AIC and \emph{Bayesian information criterion} \citep[BIC]{Schwarz1978}:
\begin{align*}
\mathrm{AIC}(\lambda,\alpha) &= \obs\log\left(\hat\sigma^2(\lambda ,\alpha)\right) +  2{df}(\lambda,\alpha),\\
\mathrm{BIC}(\lambda,\alpha) &= \obs\log\left(\hat\sigma^2(\lambda ,\alpha)\right) +  {{df}}(\lambda,\alpha)\log(\obs), 
\end{align*}
where $\hat\sigma^2(\lambda ,\alpha)=\obs^{-1}\sum_{i=1}^\obs\hat\varepsilon_i^2$ and $\hat\varepsilon_i$ are the residuals. ${{df}}(\lambda,\alpha)$ is the effective degrees of freedom, which is a measure of model complexity. In the linear regression model, the degrees of freedom is simply the number of regressors. \citet{Zou2007} show that the number of coefficients estimated to be non-zero, $\hat{s}$, is an unbiased and consistent estimate of ${{df}}(\lambda)$ for the lasso ($\alpha=1$). More generally, the degrees of freedom of the elastic net can be calculated as the trace of the projection matrix, i.e., 
 \[ \widehat{{df}}(\lambda,\alpha)=\textrm{tr}(\bm{X}_{\hat{\Omega}}(\bm{X}_{\hat{\Omega}}'\bm{X}_{\hat{\Omega}}+\lambda(1-\alpha)\bm{\Psi})^{-1}\bm{X}_{\hat{\Omega}}').\]
where $\bm{X}_{\hat{\Omega}}$ is the $n \times \hat{s}$ matrix of selected regressors.
The unbiased estimator of the degrees of freedom provides a justification for using the classical AIC and BIC to select tuning parameters \citep{Zou2007}.

The BIC is known to be model selection consistent if the true model is among the candidate models, whereas AIC is inconsistent. Clearly, the assumption that the true model is among the candidates is strong; even the existence of the `true model' may be problematic, so that loss efficiency may become a desirable second-best. The AIC is, in contrast to BIC, loss efficient. \citet{Yang2005} shows that the differences between AIC-type information criteria and BIC are fundamental; a consistent model selection method, such as the BIC, cannot be loss efficient, and vice versa. 
\citet{Zhang2010} confirm this relation in the context of penalized regression. 

Both AIC and BIC are not suitable in the large-$p$-small-$\obs$ context where they tend to select too many variables (see Monte Carlo simulations in Section~\ref{sec:mc}). It is well known that the AIC is biased in small samples, which motivated the bias-corrected AIC \citep{Sugiura1978,Hurvich1989},
\[ \mathrm{AIC}_c(\lambda,\alpha) = \obs\log\left(\hat\sigma^2(\lambda ,\alpha)\right) +  2{df}(\lambda,\alpha) \frac{\obs}{\obs-{df}(\lambda,\alpha)}. \] 
The bias can be severe if $df$ is large relative to $\obs$, and thus the AIC$_c$ should be favoured when $n$ is small or with high-dimensional data.

The BIC relies on the assumption that each model has the same prior probability. This assumptions seems reasonable when the researcher has no prior knowledge; yet, it contradicts the principle of parsimony and becomes problematic if $p$ is large. To see why, consider the case where $p=1000$ \citep[following][]{Chen2008}: There are $1000$ models for which one parameter is non-zero ($s=1$), while there are $1000\times 999/2$ models for which $s=2$. Thus, the prior probability of $s=2$ is larger than the prior probability of $s=1$ by a factor of $999/2$. More generally, since the prior probability that $s=j$ is larger than the probability that $s=j-1$ (up to the point where $j=p/2$), the BIC is likely to over-select variables. To address this shortcoming, \citet{Chen2008} introduce the Extended BIC, defined as
\[ \mathrm{EBIC}_\xi(\lambda,\alpha) = \obs\log\left(\hat\sigma^2(\lambda ,\alpha)\right) +  {{df}}(\lambda,\alpha)\log(\obs)+2\xi{{df}}(\lambda,\alpha) \log(p), \]
which imposes an additional penalty on the size of the model. The prior distribution is chosen such that the probability of a model with dimension $j$ is inversely proportional to the total number of models for which $s=j$. The additional parameter, $\xi\in[0,1]$, controls the size of the additional penalty.\footnote{We follow  \citet[][p.~768]{Chen2008} and use $\xi=1-\log(\obs)/(2\log(p))$ as the default choice.
An upper and lower threshold is applied to ensure that $\xi$ lies in the [0,1] interval.} \citet{Chen2008} show in simulation studies that the EBIC$_\xi$ outperforms the traditional BIC, which exhibits a higher false discovery rate when $p$ is large relative to $n$.%


\section{Tuning parameter selection using cross-validation}\label{sec:crossvalidation}
The aim of cross-validation is to directly assess the performance of a model on unseen data. To this end, the data is repeatedly divided into a \emph{training} and a \emph{validation data set}. The models are fit to the training data and the validation data is used to assess the predictive performance. In the context of regularized regression, cross-validation can be used to select the tuning parameters that yield the best performance, e.g., the best \emph{out-of-sample} mean squared prediction error. A wide range of methods for cross-validation are available. For an extensive review, we recommend \citet{Arlot2010}. The most popular method is $K$-fold cross-validation, which we introduce in Section~\ref{sec:kfoldcv}. In Section~\ref{sec:cvtime}, we discuss methods for cross-validation in the time-series setting.


\subsection{K-fold cross-validation}\label{sec:kfoldcv}
\begin{figure}[H]
\centering
\sjversion{\includegraphics[width=.85\linewidth]{cv_groups_mono.pdf}}{\includegraphics[width=.85\linewidth]{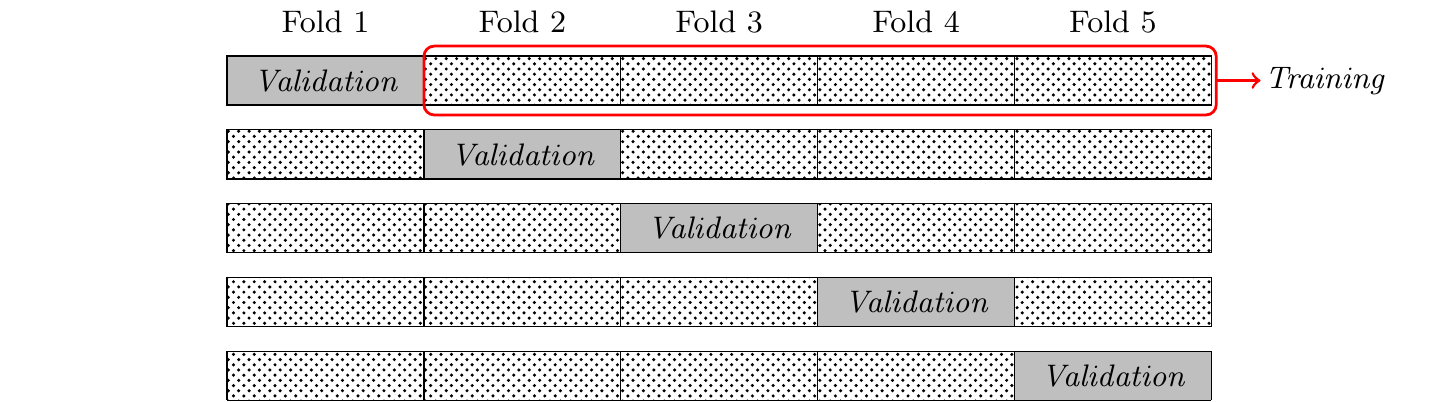}}
\caption{Data partition for 5-fold cross-validation. Each row corresponds to one step and each column to one data partition (`fold'). In the first step, fold~1 constitutes the validation data and folds~2-5 are the training data.}\label{fig:cv_groups}
\end{figure}

For $K$-fold cross-validation, proposed by \citet{Geisser1975}, the data is split into $K$ groups, referred to as folds, of approximately equal size. Let $\mathcal{K}_k$ denote the set of observations in the $k$th fold, and let $\obs_k$ be the size of data partition $k$ for $k=1,...,K$. In the $k$th step, the $k$th fold is treated as the validation data set and the remaining $K-1$ folds constitute the training data set.  The model is fit to the training data for a given value of $\lambda$ and $\alpha$.  The resulting estimate, which is based on all the data except the observations in fold $k$, is  $\bm{\hat\beta}_k(\lambda,\alpha)$. The procedure is repeated for each fold, as illustrated in Figure~\ref{fig:cv_groups}, so that every data point is used for validation once. The mean squared prediction error for each fold is computed as
    \[\textrm{MSPE}_k(\lambda,\alpha)=\frac{1}{\obs_k}\sum_{i \in \mathcal{K}_k}\left(y_i - \bm{x}_i'\bm{\hat\beta}_k(\lambda,\alpha)\right)^2.\]
    
The $K$-fold cross-validation estimate of the MSPE, which serves as a measure of prediction performance, is
    \[\hat{\mathcal{L}}^{CV}(\lambda,\alpha)=\frac{1}{K}\sum_{k=1}^K\textrm{MSPE}_k(\lambda,\alpha).\]
This suggests selecting $\lambda$ and $\alpha$ as the values that minimize $\hat{\mathcal{L}}^{CV}(\lambda,\alpha)$. An alternative common rule is to use the largest value of $\lambda$ that is within one standard deviation of the minimum, which leads to a more parsimonious model.

Cross-validation can be computationally expensive. It is necessary to compute $\hat{\mathcal{L}}^{CV}$ for each value of $\lambda$ on a grid if $\alpha$ is fixed (e.g.\ when using the lasso) or, in the case of the elastic net, for each combination of values of $\lambda$ and $\alpha$ on a two-dimensional grid. In addition, the model must be estimated $K$ times at each grid point, such that the computational cost is approximately proportional to $K$.%
\footnote{An exception is the special case of leave-one-out cross-validation, where $K=n$. The advantage of LOO cross-validation for linear models is that there is a closed-form expression for the MSPE, meaning that the model needs to be estimated only once instead of $n$ times.}

Standardization adds another layer of computational cost to $K$-fold cross validation. An important principle in cross-validation is that the training data set should not contain information from the validation dataset. This mimics the real-world situation where out-of-sample predictions are made not knowing what the true response is. The principle applies not only to individual observations, 
but also to data transformations such as mean-centering and standardization. Specifically, data transformations applied to the training data should not use information from the validation data or full dataset. Mean-centering and standardization using sample means and sample standard deviations for the full sample would violate this principle. Instead, when in each step the model is fit to the training data for a given $\lambda$ and $\alpha$, the training dataset must be re-centered and re-standardized, or, if standardization is built into the penalty loadings, the $\hat\psi_j$ must be recalculated based on the training dataset.

The choice of $K$ is not only a practical problem; it also has theoretical implications. The variance of $\hat{\mathcal{L}}^{CV}$ decreases with $K$, and is minimal (for linear regression) if $K=n$, which is referred to as leave-one-out or LOO CV. Similarly, the bias decreases with the size of the training data set. Given computational contraints, $K$ between 5 and 10 are often recommended, arguing that the performance of CV rarely improves for $K$ larger than 10 \citep{Hastie2009, Arlot2010}. 

If the aim of the researcher's analysis is model identification rather than prediction, the theory requires training data to be `small' and the evaluation sample to be close to $n$ \citep{Shao1993,Shao1997}. The reason is that more data is required to evaluate which model is the `correct' one rather than to decrease bias and variance. This is referred to as \emph{cross-validation paradox} \citep{Yang2006}. However, since $K$-fold cross-validation sets the size of the training sample to approximately $n/K$, $K$-fold CV is necessarily ill-suited for selecting the true model.  


\subsection{Cross-validation with time-series data}\label{sec:cvtime}
Serially dependent data violate the principle that training and validation data are independent. That said, standard $K$-fold cross-validation may still be appropriate in certain circumstances. \citet{Bergmeir2018} show that $K$-fold cross-validation remains valid in the pure auto-regressive model if one is willing to assume that the errors are uncorrelated.
A useful implication is that $K$-fold cross-validation can be used on overfit auto-regressive models that are not otherwise badly misspecified, since such models have uncorrelated errors.


Rolling $h$-step ahead CV is an intuitively appealing approach that directly incorporates the ordered nature of time series-data \citep{Hyndman2018}.\footnote{Another approach is a variation of LOO cross-validation known as \emph{$h$-block cross-validation} \citep{Burman1994}, which omits $h$ observations between training and validation data.} The procedure builds on repeated $h$-step ahead forecasts. The procedure is implemented in \lassopack and illustrated in Figure~\ref{tab:rollingcv1}-\ref{tab:rollingcv2}. 

\begin{figure}[H]
    \centering\footnotesize
    \begin{minipage}{.45\linewidth}\centering
    \begin{tabular}{lc|ccccc|}
          &  \multicolumn{1}{c}{}    &  \multicolumn{5}{c}{Step} \\ 
          &  \multicolumn{1}{c}{}    &  {\scriptsize 1} & {\scriptsize 2} & {\scriptsize 3} & {\scriptsize 4} & \multicolumn{1}{c}{\scriptsize 5}  \\ 
    \cline{3-7}
          &   {\scriptsize   1}&  $T$     & $T$     & $T$     & $T$       & $T$  \bigstrut[t]\\ 
          &   {\scriptsize   2}&  $T$     & $T$     & $T$     & $T$       & $T$ \\
          &   {\scriptsize   3}&  $T$     & $T$     & $T$     & $T$       & $T$ \\
      $t$ &   {\scriptsize   4}&  $V$     & $T$     & $T$     & $T$       & $T$ \\
          &   {\scriptsize   5}&  $\cdot$ & $V$     & $T$     & $T$       & $T$ \\
          &   {\scriptsize   6}&  $\cdot$ & $\cdot$ & $V$     & $T$       & $T$ \\
          &   {\scriptsize   7}&  $\cdot$ & $\cdot$ & $\cdot$ & $V$       & $T$ \\
          &   {\scriptsize   8}&  $\cdot$ & $\cdot$ & $\cdot$ & $\cdot$   & $V$ \\
    \cline{3-7}
    \end{tabular}
    \subcaption{$h=1$, expanding window}
    \end{minipage}\hfill%
    \begin{minipage}{.55\linewidth}\centering
    \begin{tabular}{lc|ccccc|}
          &  \multicolumn{1}{c}{}    &  \multicolumn{5}{c}{Step} \\ 
          &  \multicolumn{1}{c}{}    &  {\scriptsize 1} & {\scriptsize 2} & {\scriptsize 3} & {\scriptsize 4} & \multicolumn{1}{c}{\scriptsize 5}  \\ 
    \cline{3-7}
          &   {\scriptsize   1}&  $T$     & $T$     & $T$     & $T$       & $T$  \bigstrut[t]\\ 
          &   {\scriptsize   2}&  $T$     & $T$     & $T$     & $T$       & $T$ \\
          &   {\scriptsize   3}&  $T$     & $T$     & $T$     & $T$       & $T$ \\
      $t$ &   {\scriptsize   4}&  $\cdot$ & $T$     & $T$     & $T$       & $T$ \\
          &   {\scriptsize   5}&  $V$     & $\cdot$     & $T$     & $T$       & $T$ \\
          &   {\scriptsize   6}&  $\cdot$ & $V$ & $\cdot$     & $T$       & $T$ \\
          &   {\scriptsize   7}&  $\cdot$ & $\cdot$ & $V$ & $\cdot$       & $T$ \\
          &   {\scriptsize   8}&  $\cdot$ & $\cdot$ & $\cdot$ & $V$   & $\cdot$ \\
          &   {\scriptsize   9}&  $\cdot$ & $\cdot$ & $\cdot$ & $\cdot$   & $V$ \\
    \cline{3-7}
    \end{tabular}
    \subcaption{$h=2$, expanding window}
    \end{minipage}\\[.2cm]
    \caption{Rolling $h$-step ahead cross-validation with expanding training window. `$T$' and `$V$' denote that the observation is included in the training and validation sample, respectively. A dot (`.') indicates that an observation is excluded from both training and validation data.}
    \label{tab:rollingcv1}
\end{figure} 

\begin{figure}
    \centering\footnotesize
    \begin{minipage}{.45\linewidth}\centering
    \begin{tabular}{lc|ccccc|}
          &  \multicolumn{1}{c}{}    &  \multicolumn{5}{c}{Step} \\ 
          &  \multicolumn{1}{c}{}    &  {\scriptsize 1} & {\scriptsize 2} & {\scriptsize 3} & {\scriptsize 4} & \multicolumn{1}{c}{\scriptsize 5}  \\ 
    \cline{3-7}
          &   {\scriptsize   1}&  $T$     & $\cdot$     & $\cdot$     &$\cdot$       & $\cdot$  \bigstrut[t]\\ 
          &   {\scriptsize   2}&  $T$     & $T$     & $\cdot$     & $\cdot$      & $\cdot$ \\
          &   {\scriptsize   3}&  $T$     & $T$     & $T$     & $\cdot$       & $\cdot$ \\
      $t$ &   {\scriptsize   4}&  $V$     & $T$     & $T$     & $T$       & $\cdot$ \\
          &   {\scriptsize   5}&  $\cdot$ & $V$     & $T$     & $T$       & $T$ \\
          &   {\scriptsize   6}&  $\cdot$ & $\cdot$ & $V$     & $T$       & $T$ \\
          &   {\scriptsize   7}&  $\cdot$ & $\cdot$ & $\cdot$ & $V$       & $T$ \\
          &   {\scriptsize   8}&  $\cdot$ & $\cdot$ & $\cdot$ & $\cdot$   & $V$ \\
    \cline{3-7}
    \end{tabular}
    \subcaption{$h=1$, fixed window}
    \end{minipage}\hfill%
    \begin{minipage}{.55\linewidth}\centering
    \begin{tabular}{lc|ccccc|}
          &  \multicolumn{1}{c}{}    &  \multicolumn{5}{c}{Step} \\ 
          &  \multicolumn{1}{c}{}    &  {\scriptsize 1} & {\scriptsize 2} & {\scriptsize 3} & {\scriptsize 4} & \multicolumn{1}{c}{\scriptsize 5}  \\ 
    \cline{3-7}
          &   {\scriptsize   1}&  $T$     & $\cdot$     & $\cdot$     & $\cdot$       &$\cdot$  \bigstrut[t]\\ 
          &   {\scriptsize   2}&  $T$     & $T$     & $\cdot$     & $\cdot$       & $\cdot$\\
          &   {\scriptsize   3}&  $T$     & $T$     & $T$     & $\cdot$       & $\cdot$ \\
      $t$ &   {\scriptsize   4}&  $\cdot$ & $T$     & $T$     & $T$       & $\cdot$ \\
          &   {\scriptsize   5}&  $V$     & $\cdot$     & $T$     & $T$       & $T$ \\
          &   {\scriptsize   6}&  $\cdot$ & $V$ & $\cdot$     & $T$       & $T$ \\
          &   {\scriptsize   7}&  $\cdot$ & $\cdot$ & $V$ & $\cdot$       & $T$ \\
          &   {\scriptsize   8}&  $\cdot$ & $\cdot$ & $\cdot$ & $V$   & $\cdot$ \\
          &   {\scriptsize   9}&  $\cdot$ & $\cdot$ & $\cdot$ & $\cdot$   & $V$ \\
    \cline{3-7}
    \end{tabular}
    \subcaption{$h=2$, fixed window}
    \end{minipage}
    \caption{Rolling $h$-step ahead cross-validation with fixed training window.}
    \label{tab:rollingcv2}
\end{figure}

Figure~\ref{tab:rollingcv1}(a) corresponds to the default case of 1-step ahead cross-validation. `$T$' denotes the observation included in the training sample and `$V$' refers to the validation sample. In the first step, observations 1 to 3 constitute the training data set and observation~4 is the validation point, whereas the remaining observations are unused as indicated by a dot (`.'). Figure~\ref{tab:rollingcv1}(b) illustrates the case of 2-step ahead cross-validation. In  both cases, the training window expands incrementally, whereas Table~\ref{tab:rollingcv2} displays rolling CV with a fixed estimation window.


\subsection{Comparison with information criteria}
Since information-based approaches and cross-validation share the aim of model selection, one might expect that the two methods share some theoretical properties. Indeed, AIC and LOO-CV are asymptotically equivalent, as shown by \citet{Stone1977} for fixed $p$. Since information criteria only require the model to be estimated once, they are computationally much more attractive, which might suggest that information criteria are superior in practice. However, an advantage of CV is its flexibility and that it adapts better to situations where the assumptions underlying information criteria, e.g.\ homoskedasticity, are not satisfied \citep{Arlot2010}. If the aim of the analysis is identifying the true model, BIC and EBIC provide a better choice than $K$-fold cross-validation, as there are strong but well-understood conditions under which BIC and EBIC are model selection consistent.

\section{Rigorous penalization}\label{sec:rigorous}
This section introduces the `rigorous' approach to penalization. Following \citet{hdm2016}, we use the term `rigorous' to emphasize that the framework is grounded in theory. In particular, the penalization parameters are chosen to guarantee consistent prediction and parameter estimation.
Rigorous penalization is of special interest, as it provides the basis for methods to facilitate causal inference in the presence of many instruments and/or many control variables; these methods are the IV-Lasso \citep{Belloni2012}, the post-double-selection (PDS) estimator \citep{Belloni2014b} and the post-regularization estimator (CHS) \citep{Chernozhukov2015}; all of which are implemented in our sister package \pdslasso \citep{pdslasso2018}.  

We discuss the conditions required to derive theoretical results for the lasso in Section~\ref{sec:lasso_theory}. Sections~\ref{sec:rigorous_iid}-\ref{sec:rigorous_xdep} present feasible algorithms for optimal penalization choices for the lasso and square-root lasso under i.i.d., heteroskedastic and cluster-dependent errors. Section \ref{sec:rigorous_sign} presents a related test for joint significance testing.

\subsection{Theory of the lasso}\label{sec:lasso_theory}
There are three main conditions required to guarantee that the lasso is consistent in terms of prediction and parameter estimation.\footnote{For a more detailed treatment, we recommend \citet[Ch.~11]{Hastie2015} and \citet{Buhlmann2011}.} The first condition relates to sparsity. Sparsity is an attractive assumption in settings where we have a large set of potentially relevant regressors, or consider various different model specifications, but assume that only one true model exists which includes a small number of regressors. We have introduced \emph{exact sparsity} in Section~2, but the assumption is stronger than needed. For example, some true coefficients may be non-zero, but small in absolute size, in which case it might be preferable to omit them. For this reason, we use a weaker assumption:

\paragraph{Approximate sparsity.} \citet{Belloni2012} consider the \emph{approximate sparse model (ASM)},
\begin{equation}
 y_i = f(\bm{w}_i) + \varepsilon_i = \bm{x}_i'\bm{\beta}_0 + r_i + \varepsilon_i. 
\end{equation}
The elementary regressors $\bm{w}_i$ are linked to the dependent variable through the unknown and possibly non-linear function $f(\cdot)$. The aim of the lasso (and square-root lasso) estimation is to approximate $f(\bm{w}_i)$ using the target parameter vector $\bm{\beta}_0$ and the transformations $\bm{x}_i:=P(\bm{w}_i)$, where $P(\cdot)$ denotes a dictionary of transformations. The vector $\bm{x}_i$ may be large relative to the sample size, either because $\bm{w}_i$ itself is high-dimensional and $\bm{x}_i:=\bm{w}_i$, or because a large number of transformations such as dummies, polynomials, interactions are considered to approximate $f(\bm{w}_i)$. 

The assumption of approximate sparsity requires the existence of a target vector $\bm{\beta}_0$ which ensures that $f(\bm{w}_i)$ can be approximated sufficiently well, while using only a small number of non-zero coefficients. Specifically, the target vector $\bm{\beta}_0$ and the sparsity index $s$ are assumed to meet the condition
\[ \normline{\bm{\beta}_0}_0:=s \ll n \qquad \mathrm{with} \quad\frac{s^2\log^2(p\vee n)}{n}\rightarrow 0,\] 
and the resulting approximation error $r_i=f(\bm{w}_i)-\bm{x}_i'\bm{\beta}_0$ is bounded such that
\begin{equation}\sqrt{\frac{1}{n}\sum_{i=1}^nr_i^2}\leq C \sqrt{\frac{s}{n}},\label{eq:approxerror}\end{equation}
where $C$ is a positive constant. To emphasize the distinction between approximate and exact sparsity, consider the special case where $f(\bm{w}_i)$ is linear with $f(\bm{w}_i)=\bm{x}_i'\bm{\beta}^\star$, but where the true parameter vector $\bm{\beta}^\star$ violates exact sparsity so that $\normline{\bm{\beta}^\star}_0 > n$. If $\bm{\beta}^\star$ has many elements that are negligible in absolute size, we might still be able to approximate $\bm{\beta}^\star$ using the sparse target vector $\bm{\beta}_0$ as long as  $r_i=\bm{x}_i'(\bm{\beta}^\star-\bm{\beta}_0)$ is sufficiently small as specified in~\eqref{eq:approxerror}.

\paragraph{Restricted sparse eigenvalue condition.} The second condition relates to the Gram matrix, $n^{-1}\bm{X}'\bm{X}$. In the high-dimensional setting where $p$ is larger than $n$, the Gram matrix is necessarily rank-deficient and the minimum (unrestricted) eigenvalue is zero, i.e., \[ \min_{\bm{\delta}\neq 0} \frac{\norm{\bm{X}\bm{\delta}}_2}{\sqrt{n}\norm{\bm{\delta}}_2}=0.\]
Thus, to accommodate large $p$, the full rank condition of OLS needs to be replaced by a weaker condition. While the full rank condition cannot hold for the full Gram matrix if $p>n$, we can plausibly assume that sub-matrices of size $m$ are well-behaved. This is in fact the \emph{restricted sparse eigenvalue (RSEC)} condition of \citet{Belloni2012}.
The RSEC formally states that the minimum sparse eigenvalues
\[ \phi_{\min}(m) = \min_{1\leq\normline{\bm\delta}_0 \leq m} \frac{\bm\delta'\bm{X}'\bm{X}\bm\delta}{\normline{\bm\delta}^2_2} \qquad \mathrm{and} \qquad \phi_{\max}(m) = \max_{1\leq\normline{\bm\delta}_0 \leq m} \frac{\bm\delta'\bm{X}'\bm{X}\bm\delta}{\normline{\bm\delta}^2_2}\]
are bounded away from zero and from above. The requirement $\phi_{\min}(m)>0$ implies that all sub-matrices of size $m$ have to be positive definite.\footnote{The RSEC is stronger than required for the lasso. For example, \citet{Bickel2009} introduce the \emph{restricted eigenvalue condition (REC)}. However, here we only present the RSEC which implies the REC and is sufficient for both lasso and post-lasso. Different variants of the REC and RSEC have been proposed in the literature; for an overview see \citet{Buhlmann2011}.}

\paragraph{Regularization event.} The third central condition concerns the choice of the penalty level $\lambda$ and the predictor-specific penalty loadings $\psi_j$. The idea is to select the penalty parameters to control the random part of the problem in the sense that
\begin{equation} 
\frac{\lambda}{n} \geq c\max_{1\leq j\leq p} \big| \psi_j^{-1} S_j\big| \quad\textnormal{where}\quad S_j = \frac{2}{n }\sum_{i=1}^n x_{ij}\varepsilon_i
\label{eq:lambdaevent}\end{equation}
with high probability. Here, $c>1$ is a constant slack parameter and $S_j$ is the $j$th element of the score vector $\bm{S} = \nabla \hat{Q}(\bm{\beta})$, the gradient of $\hat{Q}$ at the true value $\bm{\beta}$. The score vector summarizes the noise associated with the estimation problem.%

Denote by $\Lambda = n \max_j | \psi_j^{-1} S_j |$ the maximal element of the score vector scaled by $n$ and $\psi_j$, and denote by $q_\Lambda(\cdot)$ the quantile function for $\Lambda$.\footnote{That is, the probability that $\Lambda$ is at most $a$ is $q_\Lambda(a)$.}
In the rigorous lasso, we choose the penalty parameters $\lambda$ and $\psi_j$ and confidence level $\gamma$ so that
\begin{equation}
     \lambda \geq c q_\Lambda(1-\gamma)
\label{eq:lambda_rule}
\end{equation}

A simple example illustrates the intuition behind this approach. 
Consider the case where the true model has $\beta_j=0$ for $j=1,\ldots,p$, i.e., none of the regressors appear in the true model.
It can be shown that for the lasso to select no variables, the penalty parameters $\lambda$ and $\psi_j$ need to satisfy $\lambda \geq 2 \max_j  |\sum_i \psi_j^{-1} x_{ij} y_i|$.\footnote{See, for example, \citet[Ch.~2]{Hastie2015}.}
Because none of the regressors appear in the true model, $y_i=\varepsilon_i$.
We can therefore rewrite the requirement for the lasso to correctly identify the model without regressors as $\lambda \geq 2 \max_j  |\sum_i \psi_j^{-1} x_{ij} \varepsilon_i|$, which is the regularization event in~\eqref{eq:lambdaevent}.
We want this regularization event to occur with high probability of at least $(1-\gamma)$.
If we choose values for $\lambda$ and $\psi_j$ such that  $\lambda \geq q_\Lambda(1-\gamma)$, then by the definition of a quantile function we will choose the correct model---no regressors---with probability of at least $(1-\gamma)$.
This is simply the rule in~\eqref{eq:lambda_rule}.

The chief practical problem in using the rigorous lasso is that the quantile function $q_\Lambda(\cdot)$ is unknown.
There are two approaches to addressing this problem proposed in the literature, both of which are implemented in \rlasso.
The \rlasso default is the `asymptotic' or \emph{X-independent} approach: theoretically grounded and feasible penalty level and loadings are used that guarantee that \eqref{eq:lambdaevent} holds asymptotically, as $n\rightarrow\infty$ and $\gamma\rightarrow 0$.
The \emph{X-independent} penalty level choice can be interpreted as an asymptotic upper bound on the quantile function $q_\Lambda(.)$.
In the `exact' or \emph{X-dependent} approach, the quantile function $q_\Lambda(.)$ is directly estimated by simulating the distribution of $q_\Lambda(1-\gamma|\bm{X})$, the $(1-\gamma)$-quantile of $\Lambda$ conditional on the observed regressors $\bm{X}$. 
We first focus on the \emph{X-independent approach}, and introduce the \emph{X-dependent approach} in Section~\ref{sec:rigorous_xdep}.

\subsection{Rigorous lasso} \label{sec:rigorous_iid}
\citet{Belloni2012} show---using moderate deviation theory of self-normalized sums from \citet{Jing2003}---that the regularization event in~\eqref{eq:lambdaevent} holds asymptotically, i.e.,
\begin{equation} \mathrm{P}\left( \max_{1\leq j\leq p}   c\, \big| S_j \big|\leq\frac{\lambda \psi_j}{n} \right) \rightarrow 1 ~~\textrm{as}~~n\rightarrow\infty, ~ \gamma\rightarrow 0.\label{eq:prob_reg_event}\end{equation}
if the penalty levels and loadings are set to 
 \begin{equation}
 \begin{array}{lll}
 \textrm{homoskedasticity:} & \lambda =2c\sigma \sqrt{n}\Phi^{-1}(1-\gamma/(2p)), &\quad\psi_j = \sqrt{\dfrac{1}{n}\sum_i x_{ij}^2} ,  \\  
 \textrm{heteroskedasticity:} & \lambda = 2c\sqrt{n}\Phi^{-1}(1-\gamma/(2p)), &\quad \psi_j = \sqrt{\dfrac{1}{n}\sum_i x_{ij}^2\varepsilon_i^2},
 \end{array}
 \label{eq:idealpenalty} 
 \end{equation}
under homoskedasticity and heteroskedasticity, respectively. 
$c$ is the slack parameter from above and the significance level $\gamma$ is required to converge towards 0.  \rlasso uses $c=1.1$ and $\gamma=0.1/\log(n)$ as defaults.%
\footnote{The parameters $c$ and $\gamma$ can be controlled using the options \texttt{c({\it real})} and \texttt{gamma({\it real})}.
Note that we need to choose $c$ greater than 1 for the regularization event to hold asymptotically, but not too high as the shrinkage bias is increasing in $c$.}\textsuperscript{,}%
\footnote{
An alternative \emph{X-independent} choice is to set $\lambda = 2c\sigma \sqrt{2 n \log(2p/\gamma)}$.
Since $\sqrt{n}\Phi^{-1}(1-\gamma/(2p)) \leq \sqrt{2n\log(2p/\gamma)}$, this will lead to a more parsimonious model, but also to a larger bias. To use the alternative \emph{X-independent}, specify the \texttt{lalt} option.
\label{fn:lalt}}

\paragraph{Homoskedasticity.} We first focus on the case of homoskedasticity.
In the rigorous lasso approach, we standardize the score.
But since $E(x_{ij}^2\varepsilon_i^2) = \sigma E(x_{ij}^2)$ under homoskedasticity, we can separate the problem into two parts: the regressor-specific penalty loadings $\psi_j = \sqrt{(1/n) \sum_i x_{ij}^2}$ standardize the regressors, and $\sigma$ moves into the overall penalty level.
In the special case where the regressors have already been standardized such that $(1/n)\sum_i x_{ij}^2=1$, the penalty loadings are $\psi_j=1$. Hence, the purpose of the regressor-specific penalty loadings in the case of homoskedasticity is to accommodate regressors with unequal variance. 

The only unobserved term is $\sigma$, which appears in the optimal penalty level $\lambda$. To estimate $\sigma$, we can use some initial set of residuals $\hat\varepsilon_{0,i}$ and calculate the initial estimate as   $\hat\sigma_0=\sqrt{(1/n)\sum_i \hat\varepsilon_{0,i}^2}$. A possible choice for the initial residuals is $\hat\varepsilon_{0,i}=y_i$ as in \citet{Belloni2012} and \citet{Belloni2014b}. \texttt{rlasso} uses the OLS residuals $\hat\varepsilon_{0,i}=y_i-\bm{x}_i[\mathcal{D}]'\hat{\bm{\beta}}_{OLS}$ where $\mathcal{D}$ is the set of 5 regressors exhibiting the highest absolute correlation with $y_i$.\footnote{This is also the default setting in \citet{hdm2016}. The number of regressors used for calculating the initial residuals can be controlled using the \texttt{corrnumber(\stint)} option, where 5 is the default and 0 corresponds to $\hat\varepsilon_{0,i}=y_i$.} The procedure is summarized in Algorithm~A:

\begin{sttech}[Algorithm~A: Estimation of penalty level under homoskedasticity.]
\begin{enumerate}[nosep] 
    \item Set $k=0$, and define the maximum number of iterations, $K$.
    Regress $y_i$ against the subset of $d$ predictors exhibiting the highest correlation coefficient with $y_i$ and compute the initial residuals as $\hat\varepsilon_{0,i}=\hat\varepsilon_{k,i}=y_i-\bm{x}_i[\mathcal{D}]'\hat{\bm{\beta}}_{OLS}$.  Calculate the homoskedastic penalty loadings in~\eqref{eq:idealpenalty}. 
    \item If $k\leq K$, compute the homoskedastic penalty level in~\eqref{eq:idealpenalty} by replacing $\sigma$ with $\hat\sigma_k=\sqrt{(1/n)\sum_i \hat\varepsilon_{k,i}^2},$  and obtain the rigorous lasso or post-lasso estimator $\hat{\bm{\beta}}_k$. Update the residuals $\hat\varepsilon_{k+1,i}=y_i-\bm{x}_i'\hat{\bm{\beta}}_k$. Set $k\leftarrow k+1$. 
    \item Repeat step 2 until $k>K$ or until convergence by updating the penalty level.
\end{enumerate}
\end{sttech}

The \rlasso default is to perform one further iteration after the initial estimate (i.e., $K=1$), which in our experience provides good performance. Both lasso and post-lasso can be used to update the residuals. \rlasso uses post-lasso to update the residuals.\footnote{The \texttt{lassopsi} option can be specified, if rigorous lasso residuals are preferred.}

\paragraph{Heteroskedasticity.} 
The \emph{X-independent} choice for the overall penalty level under heteroskedasticity is $\lambda = 2c\sqrt{n}\Phi^{-1}(1-\gamma/(2p))$. The only difference with the homoskedastic case is the absence of $\varsigma$. The variance of $\epsilon$ is now captured via the penalty loadings, which are set to 
$\psi_j = \sqrt{\frac{1}{n}\sum_i x_{ij}^2\varepsilon_i^2}$.
Hence, the penalty loadings here serve two purposes: accommodating both heteroskedasticity and regressors with uneven variance.

To help with the intuition, we consider the case where the predictors are already standardized. It is easy to see that, if the errors are homoskedastic with $\sigma=1$, the penalty loadings are (asymptotically) just $\psi_j=1$. If the data are heteroskedastic, however, the standardized penalty loading will not be $1$. In most practical settings, the usual pattern will be that $\hat\psi_j>1$ for some $j$. Intuitively, heteroskedasticity typically increases the likelihood that the term $\max_j |\sum_i x_{ij}\varepsilon_i|$ takes on extreme values, thus requiring a higher degree of penalization through the penalty loadings.%
\footnote{To get insights into the nature of heteroskedasticity, \rlasso also calculates and returns the standardized penalty loadings 
\begin{equation*}  \hat{\psi}_j^S = \hat{\phi}_j \left(\sqrt{\dfrac{1}{n}\sum_i x_{ij}^2}\sqrt{\dfrac{1}{n}\hat{\varepsilon}_i^2}\right)^{-1}, 
\end{equation*} which are stored in \texttt{e(sPsi)}.
}

The disturbances $\varepsilon_i$ are unobserved, so we obtain an initial set of penalty loadings $\hat\psi_j$ from an initial set of residuals $\hat{\varepsilon}_{0,i}$ similar to the i.i.d.\ case above. We summarize the algorithm for estimating the penalty level and loadings as follows:

\begin{sttech}[Algorithm~B: Estimation of penalty loadings under heteroskedasticity.]
\begin{enumerate}[nosep] 
    \item Set $k=0$, and define the maximum number of iterations, $K$.
    Regress $y_i$ against the subset of $d$ predictors exhibiting the highest correlation coefficient with $y_i$ and compute the initial residuals as $\hat\varepsilon_{0,i}=\hat\varepsilon_{k,i}=y_i-\bm{x}_i[\mathcal{D}]'\hat{\bm{\beta}}_{OLS}$.  Calculate the heteroskedastic penalty level $\lambda$ in~\eqref{eq:idealpenalty}.
    \item If $k\leq K$, compute the heterokedastic penalty loadings using the formula given in in~\eqref{eq:idealpenalty} by replacing $\varepsilon_i$ with $\hat\varepsilon_{k,i}$, obtain the rigorous lasso or post-lasso estimator $\hat{\bm{\beta}}_k$. Update the residuals $\hat\varepsilon_{k+1,i}=y_i-\bm{x}_i'\hat{\bm{\beta}}_k$. Set $k\leftarrow k+1$. 
    \item Repeat step 2 until $k>K$ or until convergence by updating the penalty loadings.
\end{enumerate}
\end{sttech}


\paragraph{Theoretical property.}
Under the assumptions SEC, ASM and if penalty level $\lambda$ and the penalty loadings are estimated by Algorithm~A or B, the lasso and post-lasso obey:\footnote{For the sake of brevity, we omit additional technical conditions relating to the moments of the error and the predictors. These conditions are required to make use of the moderate deviation theory of self-normalized sums \citep{Jing2003}, which is employed to relax the assumption of Gaussian errors. See condition RF in \citet{Belloni2012} and condition SM in \citet{Belloni2014b}.}
\begin{align} 
\sqrt{\frac{1}{n} \sum_{i=1}^n \left(\bm{x_i}'\bm{\hat{\beta}}-\bm{x_i}'\bm\beta\right)^2} &=O\left(\sqrt{\frac{s\log(p\vee n)}{n}}\right), \label{eq:theorem1a}\\
\normline{\bm{\hat{\beta}}-\bm\beta}_1 &=O\left(\sqrt{\frac{s^2\log(p\vee n)}{n}}\right),\label{eq:theorem1b}
\end{align}

The first relation in \eqref{eq:theorem1a} provides an asymptotic bound for the prediction error, and the second relation in \eqref{eq:theorem1b} bounds the bias in estimating the target parameter $\bm\beta$. \citet{Belloni2012} refer to the above convergence rates as \emph{near-oracle} rates. If the identity of the $s$ variables in the model were known, the prediction error would converge at the oracle rate $\sqrt{s/n}$. Thus, the logarithmic term $\log(p\vee n)$ can be interpreted as the cost of not knowing the true model. 

\subsection{Rigorous square-root lasso}\label{sec:rigorous_sqrt}
The theory of the square-root lasso is similar to the theory of the lasso  \citep{Belloni2011a,Belloni2014c}. The $j$th element of the score vector is now defined as 
\begin{equation}S_j = \frac{\frac{1}{n}\sum_{i=1}^n x_{ij}\varepsilon_i}{\left\{\frac{1}{n} \sum_{i=1}^n \varepsilon_i^2\right\}^{1/2} }.   \label{eq:sqrtlambdaevent}\end{equation}
To see why the square-root lasso is of special interest, we define the standardized errors $\nu_i$ as $\nu_i=\varepsilon_i/\sigma$. The $j$th element of the score vector becomes 
\begin{equation} S_j = \frac{\frac{1}{n}\sum_{i=1}^n x_{ij}\sigma\nu_i}{\left\{\frac{1}{n} \sum_{i=1}^n \sigma^2\nu_i^2\right\}^{1/2} } = \frac{\frac{1}{n}\sum_{i=1}^n x_{ij} \nu_i}{\left\{\frac{1}{n} \sum_{i=1}^n  \nu_i^2\right\}^{1/2} }
\end{equation}
and is thus independent of $\sigma$.
For the same reason, the optimal penalty level for the square-root lasso in the i.i.d.\ case,
\begin{equation}
\lambda = c\sqrt{n}\Phi^{-1}(1-\gamma/(2p)),
\label{eq:sqrt_lasso_lambda}
\end{equation}
is independent of the noise level $\sigma$. 

\paragraph{Homoskedasticity.}
The ideal penalty loadings under homoskedasticity for the square-root lasso are given by formula~(iv) in Table~\ref{tab:loadings}, which provides an overview of penalty loading choices.
The ideal penalty parameters are independent of the unobserved error, which is an appealing theoretical property and implies a practical advantage.
Since both $\lambda$ and $\psi_j$ can be calculated from the data, the rigorous square-root lasso is a one-step estimator under homoskedasticity.
\citet{Belloni2011a} show that the square-root lasso performs similarly to the lasso with infeasible ideal penalty loadings.

\paragraph{Heteroskedasticity.}
In the case of heteroskedasticity, the optimal square-root lasso penalty level remains \eqref{eq:sqrt_lasso_lambda}, but the penalty loadings, given by formula~(v) in Table~\ref{tab:loadings}, depend on the unobserved error and need to be estimated.
Note that the updated penalty loadings using the residuals $\hat\varepsilon_i$ employ thresholding: the penalty loadings are enforced to be greater than or equal to the loadings in the homoskedastic case.
The \rlasso default algorithm used to obtain the penalty loadings in the heteroskedastic case is analogous to Algorithm~B.\footnote{The \rlasso default for the square-root lasso uses a first-step set of initial residuals. The suggestion of  \cite{Belloni2014c} to use initial penalty loadings for regressor $j$ of $\hat\psi_{0,j} = \max_i |x_{ij}|$ is available using the \texttt{maxabsx} option.}
While the ideal penalty loadings are not independent of the error term if the errors are heteroskedastic, the square-root lasso may still have an advantage over the lasso, since the ideal penalty loadings are pivotal with respect to the error term up to scale, as pointed out above. 


\subsection{Rigorous penalization with panel data}\label{sec:rigorous_panel}
\citet{Kozbur2015} extend the rigorous framework to the case of clustered data, where a limited form of dependence---within-group correlation---as well as heteroskedasticity are accommodated.
They prove consistency of the rigorous lasso using this approach in the large $n$, fixed $T$ and large $n$, large $T$ settings. The authors present the approach in the context of a fixed-effects panel data model, $y_{it}=\bm{x}_{it}'\bm{\beta}+\mu_i+\varepsilon_{it}$, and apply the rigorous lasso after the within transformation to remove the fixed effects $\mu_i$. The approach extends to any clustered-type setting and to balanced and unbalanced panels. For convenience we ignore the fixed effects and write the model as a balanced panel:
\begin{equation}
 y_{it}=\bm{x}_{it}'\bm{\beta}+\varepsilon_{it} \quad \quad i=1,\ldots,n, ~ t=1,\ldots,T
\end{equation}

The intuition behind the \citet{Kozbur2015} approach is similar to that behind the clustered standard errors reported by various Stata estimation commands: observations within clusters are aggregated to create `super-observations' which are assumed independent, and these super-observations are treated similarly to cross-sectional observations in the non-clustered case.
Specifically, define for the $i$th cluster and $j$th regressor the super-observation $u_{ij} := \sum_t x_{ijt}\varepsilon_{it}$. Then the penalty loadings for the clustered case are
$$\psi_j = \sqrt{\frac{1}{nT}\sum_{i=1}^n u_{ij}^2},$$ which resembles the heteroskedastic case.
The \rlasso default for the overall penalty level is the same as in the heteroskedastic case, $\lambda  = 2c\sqrt{n}\Phi^{-1}(1-\gamma/(2p))$, except that the default value for $\gamma$ is $0.1/\log(n)$, i.e., we use the number of clusters $n$ rather than the number of observations $nT$. \texttt{lassopack} also implements the rigorous square-root lasso for panel data, which uses the overall penalty in \eqref{eq:sqrt_lasso_lambda} and the penalty loadings in formula~(vi), Table~\ref{tab:loadings}.

\begin{table}
    \centering\small
\newcommand\T{\rule{0pt}{7ex}}       
\newcommand\B{\rule[-5ex]{0pt}{0ex}} 
    \begin{tabular}{l|cc|cc}\hline\hline
       & \multicolumn{2}{l|}{lasso} & \multicolumn{2}{l}{square-root lasso}  \\ \hline
      homoskedasticity      & (i) &$\displaystyle\sqrt{\frac{1}{n}\sum_{i=1}^n x_{ij}^2}$ & (iv)& $\displaystyle\sqrt{\frac{1}{n}\sum_{i=1}^n x_{ij}^2}$\T\B\\[.4cm]
      \hline
      heteroskedasticity    & (ii)& $\displaystyle\sqrt{\frac{1}{n}\sum_{i=1}^n x_{ij}^2\varepsilon_i^2}$ & (v) &$\displaystyle \sqrt{\frac{1}{n}\sum_{i=1}^n x_{ij}^2} ~ \vee ~ \sqrt{\frac{\sum_{i=1}^n x_{ij}^2\varepsilon_i^2}{\sum_{i=1}^n \varepsilon_i^2}}$\T\B\\
      \hline
      cluster-dependence      & (iii) &$\displaystyle\sqrt{\frac{1}{nT}\sum_{i=1}^n u_{ij}^2}$  & (vi)& $\displaystyle \sqrt{\frac{1}{nT}\sum_{i=1}^n \sum_t^T x_{ijt}^2} ~ \vee ~  \sqrt{\frac{\sum_{i=1}^n u_{ij}^2}{\sum_{i=1}^n \sum_{t=1}^T \varepsilon_{it}^2}}$ \T\B\\[.4cm]    \hline\hline
      \multicolumn{5}{l}{\emph{Note:} Formulas (iii) and (vi) use the notation  $u_{ij} = \sum_t x_{itj} \varepsilon_{it}$.}\\
    \end{tabular}
    \caption{Ideal penalty loadings for the lasso and square-root lasso under homoskedasticity, heteroskedasticity and cluster-dependence.}
    \label{tab:loadings}
\end{table}


\subsection{X-dependent lambda}\label{sec:rigorous_xdep}
There is an alternative, sharper choice for the overall penalty level, referred to as the \emph{X-dependent} penalty.
Recall that the asymptotic, \emph{X-independent} choice in \eqref{eq:idealpenalty} can be interpreted as an asymptotic upper bound on the quantile function of $\Lambda$, which is the scaled maximum value of the score vector. 
Instead of using the asymptotic choice, we can estimate by simulation the distribution of $\Lambda$ conditional on the observed $\bm{X}$, and use this simulated distribution to obtain the quantile $q_\Lambda(1-\gamma|\bm{X})$. 

\begin{table}
    \centering\small
\newcommand\T{\rule{0pt}{7ex}}       
\newcommand\B{\rule[-5ex]{0pt}{0ex}} 
    \begin{tabular}{l|c@{\hspace{1.2\tabcolsep}}c|c@{\hspace{1.2\tabcolsep}}c}\hline\hline
       & \multicolumn{2}{l|}{lasso} & \multicolumn{2}{l}{square-root lasso}  \\ \hline
      homoskedasticity      & (i) &$\displaystyle 2 \hat \sigma \max_{1 \leq j \leq p} \left| \sum_{i=1}^n \psi_j^{-1} x_{ij} g_i \right|$ & (iv)& $\displaystyle   \frac{1}{\hat\sigma_g} \max_{1 \leq j \leq p} \left| \sum_{i=1}^n \psi_j^{-1} x_{ij} g_i \right|$\T\B\\[.4cm]
      \hline
      heteroskedasticity    & (ii)& $\displaystyle 2 \max_{1 \leq j \leq p} \left| \sum_{i=1}^n \psi_j^{-1} x_{ij} \hat\varepsilon_i g_i \right|$ & (v) &$\displaystyle   \frac{1}{\hat\sigma_g} \max_{1 \leq j \leq p} \left| \sum_{i=1}^n \psi_j^{-1} x_{ij} \hat\varepsilon_i g_i \right|$\T\B\\
      \hline
      cluster-dependence      & (iii) &$\displaystyle  2 \max_{1 \leq j \leq p} \left| \sum_{i=1}^n \psi_j^{-1} \hat u_{ij} g_i \right|$  & (vi)& $\displaystyle  \frac{1}{\hat\sigma_g} \max_{1 \leq j \leq p} \left| \sum_{i=1}^n \psi_j^{-1} \hat u_{ij} g_i \right|$ \T\B\\[.4cm]    \hline\hline
      \multicolumn{5}{l}{\rule{0pt}{4ex}\parbox{.9\textwidth}{\emph{Note:} $g_i$ is an \emph{i.i.d.}\ standard normal variate drawn independently of the data; $\hat\sigma_g = \frac{1}{n} \sum_i g_i^2$. Formulas (iii) and (vi) use the notation  $\hat{u}_{ij} = \sum_t x_{itj} \hat{\varepsilon}_{it}$.}}\\    \end{tabular}
    \caption{Definition of $W$ statistic for the simulation of the distribution of $\Lambda$ for the lasso and square-root lasso under homoskedasticity, heteroskedasticity and cluster-dependence.}
    \label{tab:simulation}
\end{table}

In the case of estimation by the lasso under homoskedasticity, we simulate the distribution of $\Lambda$ using the statistic $W$, defined as 
\[ W= 2 \hat \sigma \max_{1 \leq j \leq p} \left| \sum_{i=1}^n \psi_j^{-1} x_{ij} g_i \right|, \] where $\psi_j$ is the penalty loading for the homoskedastic case, $\hat \sigma$ is an estimate of the error variance using some previously-obtained residuals, and $g_i$ is an \emph{i.i.d.}\ standard normal variate drawn independently of the data.
The \emph{X-dependent} penalty choice is sharper and adapts to correlation in the regressor matrix \citep{Belloni2011}.

Under heteroskedasticity, the lasso \emph{X-dependent} penalty is obtained by a multiplier bootstrap procedure.
In this case the simulated statistic $W$ is defined as in formula~(ii) in Table~\ref{tab:simulation}.
The cluster-robust \emph{X-dependent} penalty is again obtained analogously to the heteroskedastic case by defining super-observations, and the statistic $W$ is defined as in formula~(iii) in Table~\ref{tab:simulation}.
Note that the standard normal variate $g_i$ varies across but not within clusters.

The \emph{X-dependent} penalties for the square-root lasso are similarly obtained from quantiles of the simulated distribution of the square-root lasso $\Lambda$,
and are given by formulas~(iv), (v) and (vi) in Table~\ref{tab:simulation} for the homoskedastic, heteroskedastic and clustered cases, respectively.%
\footnote{Note that since $g_i$ is standard normal, in practice the term $\frac{1}{\hat\sigma_g}$ that appears in the expressions for the square-root lasso $W$ will be approximately $1$.}

\subsection{Significance testing with the rigorous lasso}\label{sec:rigorous_sign}
Inference using the lasso, especially in the high-dimensional setting, is a challenging and ongoing research area of research (see Footnote~\ref{fn:inference_hdm}).
A test that has been developed and is available in \rlasso corresponds to the test for joint significance of regressors using $F$ or $\chi^2$ tests that is common in regression analysis.
Specifically, \citet{Belloni2012} suggest using the \citet[][Appendix~M]{Kato2013} sup-score test to test for the joint significance of the regressors, i.e., 
\[ H_0: \beta_1 = \ldots = \beta_p = 0. \]  
As in the preceding sections, regressors are assumed to be mean-centered and in their original units.

If the null hypothesis is correct and the rest of the model is well-specified, including the assumption that the regressors are orthogonal to the disturbance $\varepsilon_i$, then $y_i = \varepsilon_i$ and hence $E(\bm{x}_{i}\varepsilon_i) = E(\bm{x}_i y_i) = 0$.
The sup-score statistic is
\begin{equation}
S_{SS}=\sqrt{n} \max_{1 \leq j \leq p} \left| \frac{1}{n}\sum_{i=1}^n \psi_j^{-1} x_{ij} y_i \right|
\label{eq:supscore}\end{equation}
where $\psi_j = \sqrt{\frac{1}{n}\sum_{i=1}^n (x_{ij} y_i)^2}$.
Intuitively, the $\psi_j$ in \eqref{eq:supscore} plays the same role as the penalty loadings do in rigorous lasso estimation.

The $p$-value for the sup-score test is obtained by a multiplier bootstrap procedure simulating the distribution of $S_{SS}$ by the statistic $W$, defined as 
\[ W=\sqrt{n} \max_{1 \leq j \leq p} \left| \frac{1}{n}\sum_{i=1}^n \psi_j^{-1} x_{ij} y_i g_i \right|, \] where $g_i$ is an \emph{i.i.d.} standard normal variate drawn independently of the data and $\psi_j$ is defined as in \eqref{eq:supscore}.


The procedure above is valid under both homoskedasticity and heteroskedasticity, but requires independence.
A cluster-robust version of the sup-score test that accommodates within-group dependence is
\begin{equation}
   S_{SS} = \sqrt{nT} \max_{1 \leq j \leq p} \Big| \frac{1}{nT} \sum_{i=1}^n \psi_j^{-1} u_{ij}  \Big|
\end{equation}
where $u_{ij} := \sum_{t=1}^T x_{ijt}y_{it}$ and $\psi_j = \frac{1}{nT}\sum_{i=1}^n u_{ij}^2$.

The $p$-value for the cluster version of $S_{SS}$ comes from simulating
\[ W = \sqrt{nT} \max_{1 \leq j \leq p} \Big| \frac{1}{nT} \sum_{i=1}^n \psi_j^{-1} u_{ij} g_i \Big| \]
where again $g_i$ is an \emph{i.i.d.} standard normal variate drawn independently of the data. Note again that $g_i$ is invariant within clusters.

\texttt{rlasso} also reports conservative critical values for $S_{SS}$ using an asymptotically-justified upper bound: $CV = c\Phi^{-1}(1-\gamma_{SS}/(2p))$.
The default value of the slack parameter is $c=1.1$ and the default test level is $\gamma_{SS}=0.05$. These parameters can be varied using the  \texttt{c({\it real})} and \texttt{ssgamma({\it real})} options, respectively.
The simulation procedure to obtain $p$-values using the simulated statistic $W$ can be computationally intensive, so users can request reporting of only the computationally inexpensive conservative critical values by setting the number of simulated values to zero with the \texttt{ssnumsim(\it int)} option.

\section{The commands}\label{sec:commands}
The package \lassopack consists of three commands: \lassotwo, \cvlasso and \rlasso. Each command offers an alternative method for selecting the tuning parameters $\lambda$ and $\alpha$. We discuss each command in turn and present its syntax. We focus on the main syntax and the most important options. Some technical options are omitted for the sake of brevity. For a full description of syntax and options, we refer to the help files. 

\subsection{lasso2: Base command and information criteria}
The primary purpose of \lassotwo is to obtain the solution of (adaptive) lasso, elastic net and square-root lasso for a single value of $\lambda$ or a list of penalty levels, i.e., for $\lambda_1, \ldots, \lambda_r, \ldots, \lambda_R$, where $R$ is the number of penalty levels. The basic syntax of \lassotwo is as follows: 

\begin{stsyntax}
lasso2 
\depvar\
\textit{indepvars}\ 
\optif\ 
\optin\ 
\optional{,
\underbar{alp}ha({\real})\
sqrt\
\underbar{ada}ptive\
\underbar{adal}oadings({\stnumlist})\
\underbar{adat}heta({\it real})\
ols\ 
\underbar{l}ambda({\stnumlist})\
\underbar{lc}ount({\stint})\
\underbar{lminr}atio({\real})\
lmax({\real})\
\underbar{notp}en({\varlist})\
\underbar{par}tial({\varlist})\
\underbar{pload}ings({\ststring})\
\underbar{unitl}oadings\
prestd\
\underbar{stdc}oef\
fe\
\underbar{nocons}tant\
\underbar{tolo}pt({\real})\ 
\underbar{tolz}ero({\real})\ 
\underbar{maxi}ter({\stint})\ 
\underbar{plotp}ath({\method})\ 
\underbar{plotv}ar({\varlist})\ 
\underbar{ploto}pt({\ststring})\ 
\underbar{plotl}abel\ 
lic({\ststring})\ 
ic({\ststring})\ 
\texttt{\underbar{ebicx}i}(\real)\
\underbar{postres}ults%
}
\end{stsyntax}

The options \stcmd{\underbar{alp}ha({\it real})}, \stcmd{sqrt}, \stcmd{\underbar{ada}ptive} and \texttt{ols} can be used to select elastic net, square-root lasso, adaptive lasso and post-estimation OLS, respectively. The default estimator of \lassotwo is the lasso, which corresponds to \stcmd{\underbar{alp}ha(1)}.  The special case of \stcmd{alpha(0)} yields ridge regression. 

The behaviour of \lassotwo depends on whether the {\it numlist} in \stcmd{\underbar{l}ambda({\it numlist})} is of length greater than one or not.  If {\it numlist} is a list of more than one value, the solution consists of a matrix of coefficient estimates which are stored in \texttt{e(betas)}. Each row in \texttt{e(betas)} corresponds to a distinct value of $\lambda_r$ and each column to one of the predictors in \textit{indepvars}. The `path' of coefficient estimates over $\lambda_r$ can be plotted using \stcmd{\underbar{plotp}ath({\it method})}, where {\it method} controls whether the coefficient estimates are plotted against lambda (`lambda'), the natural logarithm of lambda (`lnlambda') or the $\ell_1$-norm (`norm'). If the {\it numlist} in \stcmd{\underbar{l}ambda({\it numlist})} is a scalar value, the solution is a vector of coefficient estimates which is stored in \texttt{e(b)}. The default behaviour of \lassotwo is to use a list of 100 values.

In addition to obtaining the coefficient path, \lassotwo calculates four information criteria (AIC, AIC$_c$, BIC and EBIC). These information criteria can be used for model selection by choosing the value of $\lambda_r$ that yields the lowest value for one of the four information criteria. The \stcmd{ic({\it string})} option controls which information criterion is shown in the output, where {\it string} can be replaced with `aic', `aicc', `bic or `ebic'. \stcmd{lic({\it string})} displays the estimation results corresponding to the model selected by an information criterion. It is important to note that \stcmd{lic({\it string})} will not store the results of the estimation. This has the advantage that the user can compare the results for different information criteria without the need to re-estimate the full model. To save the estimation results, the \stcmd{\underbar{postr}esults} option should be specified in combination with \stcmd{lic({\it string})}.

\subsubsection{Estimation methods}
\hangpara
\texttt{\underbar{alp}ha({\it real})}\ controls the elastic net parameter, $\alpha$, which controls the degree of $\ell_1$-norm (lasso-type) to $\ell_2$-norm
                            (ridge-type) penalization.  \texttt{alpha(1)} corresponds to the lasso (the default estimator), and
                            \texttt{alpha(0)} to ridge regression.  The {\it real} value must be in the interval [0,1].
\par\hangpara
\texttt{sqrt}\ specifies the square-root lasso estimator. Since the square-root lasso does not employ any form of $\ell_2$-penalization, the option is incompatible with \texttt{alpha({\it real})}.
\par\hangpara
\texttt{\underbar{ada}ptive}\ specifies the adaptive lasso estimator.  The penalty loading for predictor $j$ is set to
                            $|\hat\beta_{j,0}|^{-\theta}$ where $\hat\beta_{j,0}$ is the OLS estimator or univariate OLS estimator if
                            $p>n$.  $\theta$ is the adaptive exponent, and can be controlled using the \texttt{adatheta({\it real})}
                            option.
\par\hangpara                          
\texttt{\underbar{adal}oadings({\it matrix})}\  is a matrix of   alternative initial estimates, $\hat\beta_{j,0}$, used for calculating adaptive loadings.  For
                            example, this could be the vector \texttt{e(b)} from an initial \stcmd{lasso2} estimation.  The absolute value of $\hat\beta_{j,0}$
                            is raised to the power $-\theta$ (note the minus).  
\par\hangpara                            
\texttt{\underbar{adat}heta({\it real})}\    is the exponent for calculating adaptive penalty loadings. 
The default is \texttt{\underbar{adat}heta(1)}
\par\hangpara
\texttt{ols}\ specifies that post-estimation OLS estimates are displayed and
                            returned in \texttt{e(betas)} or \texttt{e(b)}.  

\subsubsection{Options relating to lambda}
\hangpara
\texttt{\underbar{l}ambda({\stnumlist})} controls the penalty level(s), $\lambda_r$, used for estimation. \stnumlist\ is a scalar value or list of values in descending order. 
Each $\lambda_r$ must be greater than zero. If not specified, the default list is used which, using Mata syntax, is defined by\\[.2cm]
\centerline{\texttt{exp(rangen(log(lmax),log(lminratio*lmax),lcount))},}\\[.2cm]
where \texttt{lcount}, \texttt{lminratio} and \texttt{lmax} are defined below, \texttt{exp()} is the exponential function, \texttt{log()} is the natural logarithm, 
and \texttt{rangen({\it a},{\it b},{\it n})} creates a column vector going from \textit{a} to \textit{b} in \textit{n}-1 steps (see the \texttt{mf\_range} help file).
Thus, the default list ranges from \texttt{lmax} to \texttt{lminratio*lmax} and \texttt{lcount} is the number of values.
The distance between each $\lambda_r$ in the sequence is the same on the logarithmic scale.
\par\hangpara
\texttt{\underbar{lc}ount({\stint})} is the number of penalty values, $R$, for which the solution is obtained. The default is \texttt{lcount(100)}.
\par\hangpara
\texttt{lmax({\real})}   is the maximum value penalty level, $\lambda_1$. By default, $\lambda_1$ is chosen as the smallest penalty level for which the model is empty. 
Suppose the regressors are mean-centered and standardized, then $\lambda_1$ is defined as 
 $\max_j \frac{2}{n\alpha}\sum_{i=1}^n |x_{ij}y_i|$
for the elastic net and
 $\max_j \frac{1}{n\alpha}\sum_{i=1}^n |x_{ij}y_i|$
for the square-root lasso \citep[see][Section~2.5]{Friedman2010}.
\par\hangpara
\texttt{\underbar{lminr}atio({\real})}  is the ratio of the minimum penalty level, $\lambda_R$, to maximum penalty level, $\lambda_1$. \real\ must be between 0 and 1. Default is \texttt{\underbar{lminr}atio(0.001)}.

\subsubsection{Information criteria}
\par\hangpara
\texttt{lic({\ststring})}   specifies that, after the first \lassotwo estimation using a list of penalty levels, the model that corresponds to the minimum information criterion will be estimated and displayer. `aic', `bic', `aicc', and `ebic' (the default) are allowed. However, the results are not stored in \texttt{e()}.
\par\hangpara
\texttt{\underbar{postr}esults} is used in combination with \texttt{lic({\it string})}. 
\texttt{\underbar{postr}esults} stores estimation results of the model selected by information criterion in \texttt{e()}.\footnote{This option was called \texttt{postest} in earlier versions of \lassopack.}
\par\hangpara
\texttt{ic({\ststring})}             controls which information criterion is shown in the output of \lassotwo when \texttt{lambda()} is a list.  'aic', 'bic', 'aicc', and
                            'ebic' (the default are allowed).  
\par\hangpara
\texttt{\underbar{ebicx}i}(\real) controls the $\xi$ parameter of the EBIC.  $\xi$ needs to lie in the [0,1] interval.   $\xi=0$ is equivalent to the BIC.  The default choice is $\xi=1-\log(\obs)/(2\log(p))$.

\subsubsection{Penalty loadings and standardisation}
\par\hangpara
\texttt{\underbar{notp}en(\varlist)} sets penalty loadings to zero for predictors in \varlist.  Unpenalized predictors are always included in the model.
\par\hangpara
\texttt{\underbar{par}tial(\varlist)}  specified that variables in \varlist\ are partialled out prior to estimation.
\par\hangpara
\texttt{\underbar{pload}ings({\stmatrix})}    is a row-vector of penalty loadings, and overrides the default standardization loadings.
The size of the vector should equal the number of predictors (excluding partialled-out variables and excluding the constant).
\par\hangpara
\texttt{\underbar{unitl}oadings}  specifies that penalty loadings be set to a vector of ones; overrides the default standardization loadings.
\par\hangpara
\texttt{prestd}   specifies that dependent variable and predictors are standardized prior to estimation rather than standardized ``on the fly'' using penalty loadings.  See  Section~\ref{sec:technical_standardization} for more details.  By default the coefficient estimates are un-standardized (i.e., returned in original units).
\par\hangpara
\texttt{stdcoef}   returns coefficients in standard deviation units, i.e., do not un-standardize.  Only
                            supported with \texttt{prestd} option.

\subsubsection{Penalty loadings and standardisation}
\par\hangpara
    \texttt{fe}                     within-transformation is applied prior to estimation. The option requires the data in memory to be \texttt{xtset}.
\par\hangpara
    \texttt{\underbar{nocons}tant}            suppress constant from estimation.  Default behaviour is to partial the constant out
                            (i.e., to center the regressors).

\subsubsection{Replay syntax}
The replay syntax of \lassotwo allows for plotting and changing display options, without the need to re-run the full model. It can also be used to estimate the model using the value of $\lambda$ selected by an information criterion. The syntax is given by:

\begin{stsyntax}
lasso2 
\optional{,
\texttt{\underbar{plotp}ath({\ststring})}  
\texttt{\underbar{plotv}ar(\varlist)} 
\texttt{\underbar{ploto}pt({\ststring})} 
\texttt{\underbar{plotl}abel} 
\texttt{\underbar{postr}esults}
\texttt{lic({\method})}  
\texttt{ic({\method})}%
}
\end{stsyntax}

\subsubsection{Prediction syntax}
\begin{stsyntax}
predict 
\opttype\
{\it newvar}\ 
\optif\ 
\optin\ 
\optional{,
\texttt{xb}  
\texttt{\underbar{r}esiduals} 
\texttt{ols} 
\texttt{\underbar{l}ambda(\real)} 
\texttt{lid(\stint)}
\texttt{\underbar{appr}ox}
\texttt{\underbar{noi}sily}  
\texttt{\underbar{postr}esults}%
}
\end{stsyntax}

\par\hangpara 
\texttt{xb}                     computes predicted values (the default).
\par\hangpara 
\texttt{\underbar{r}esiduals}              computes residuals.
\par\hangpara 
\texttt{ols}    specifies that post-estimation OLS will be used for prediction.

If the previous \lassotwo estimation uses more than one penalty level (i.e. $R>1$), the following options are applicable:
    
\par\hangpara 
\texttt{\underbar{l}ambda(\real)}  specifies that lambda value  used for prediction. 
\par\hangpara 
\texttt{lid(\stint)}  specifies the index of the lambda value used for prediction.
\par\hangpara 
\texttt{\underbar{appr}ox}  specifies that linear approximation is used instead of re-estimation.  Faster, but only exact if
coefficient path is piece-wise linear.
\par\hangpara 
\texttt{\underbar{noi}sily} prompts display of estimation output if re-estimation required.
\par\hangpara 
\texttt{\underbar{postr}esults} stores estimation results in \texttt{e()} if re-estimation is used.

\subsection{Cross-validation with cvlasso}
\cvlasso implements $K$-fold and $h$-step ahead rolling cross-validation. The syntax of \cvlasso is:

\begin{stsyntax}
cvlasso
\depvar\
\textit{indepvars}\ 
\optif\ 
\optin\ 
\optional{,
\underbar{alp}ha({\stnumlist})\ 
\underbar{alphac}ount({\stint})\ 
sqrt\ 
\underbar{ada}ptive\ 
\underbar{adal}oadings({\ststring})\ 
\underbar{adat}heta({\real})\ 
ols\ 
\underbar{l}ambda({\stnumlist})\ 
\underbar{lc}ount({stinteger})\ 
\underbar{lminr}atio({\real})\ 
lmax({\real})\ 
lopt\ 
lse\ 
\underbar{notp}en(\varlist)\ 
\underbar{par}tial(\varlist)\ 
\underbar{pload}ings({\ststring})\ 
\underbar{unitl}oadings\ 
prestd\ 
fe\ 
\underbar{nocons}tant\ 
\underbar{tolo}pt({\real})\ 
\underbar{tolz}ero({\real})\ 
\underbar{maxi}ter({\stint})\ 
\underbar{nf}olds({\stint})\ 
\underbar{foldv}ar(varname)\ 
\underbar{savef}oldvar(varname)\ 
\underbar{roll}ing\ 
h({\stint})\ 
\underbar{or}igin({\stint})\ 
\underbar{fixedw}indow\ 
seed({\stint})\ 
plotcv\ 
plotopt({\ststring})\ 
saveest({\ststring})%
}
\end{stsyntax}

The \texttt{\underbar{alp}ha()} option of \cvlasso option accepts a \stnumlist, while \lassotwo only accepts a scalar. If the \stnumlist\ is a list longer than one, \cvlasso cross-validates over $\lambda_r$ with $r=1,\ldots,R$ and $\alpha_m$ with $m=1,\ldots,M$. 

\texttt{plotcv} creates a plot of the estimated mean-squared prediction error as a function of $\lambda_r$, and \texttt{plotopt({\ststring})} can be used to pass plotting options to Stata's \texttt{line} command. 

Internally, \cvlasso calls \lassotwo repeatedly. Intermediate \lassotwo results can be stored using \texttt{saveest({\ststring})}.

\subsubsection{Options for K-fold cross-validation}
\par\hangpara
\texttt{\underbar{nf}olds(\stint)}  is      the number of folds used for $K$-fold cross-validation. The default is \texttt{\underbar{nf}olds(10)}, or $K=10$.
\par\hangpara
\texttt{\underbar{foldv}ar(\stvarname)}   can be used to specify what fold (data partition) each observation lies in. \stvarname\ is an integer variable
with values ranging from 1 to $K$.  If not specified, the fold variable is randomly generated such that each fold is of approximately equal size.
\par\hangpara
\texttt{\underbar{savef}oldvar(\stvarname)}   saves the fold variable variable in \stvarname.  
\par\hangpara
\texttt{seed(\stint)}          sets the seed for the generation of a random fold variable. 

\subsubsection{Options for h-step ahead rolling cross-validation}
\par\hangpara
\texttt{\underbar{roll}ing}  uses rolling $h$-step ahead cross-validation. The option requires the data to be \stcmd{tsset} or \stcmd{xtset}.
\par\hangpara
\texttt{h(\stint)} changes the forecasting horizon. The default is \texttt{h(1)}.
\par\hangpara
\texttt{\underbar{or}igin(\stint)} controls the number of observations in the first training dataset.
\par\hangpara
\texttt{\underbar{fixedw}indow}   ensures that the size of the training data set is constant.

\subsubsection{Options for selection of lambda}
\par\hangpara
\texttt{lopt}  specifies that, after cross-validation, \lassotwo estimates the model with the value of $\lambda_r$ that minimizes the mean-squared
                            prediction error. That is, the model is estimated with $\lambda=\hat\lambda_\texttt{lopt}$.
\par\hangpara
\texttt{lse}  specifies that, after cross-validation, \lassotwo estimates model with largest $\lambda_r$ that is within one standard
                            deviation from $\hat\lambda_\texttt{lopt}$. That is, the model is estimated with $\lambda=\hat\lambda_\texttt{lse}$.
\par\hangpara
\texttt{\underbar{postr}esults} stores the \lassotwo estimation results in \texttt{e()} (to be used in combination with \texttt{lse} or \texttt{lopt}).

\subsubsection{Replay syntax}
Similar to \lassotwo, \cvlasso also provides a replay syntax, which helps to avoid time-consuming re-estimations. The replay syntax of \cvlasso can be used for plotting and to estimate the model corresponding to $\hat\lambda_\texttt{lopt}$ or $\hat\lambda_\texttt{lse}$. The replay syntax of \cvlasso is given by:

\begin{stsyntax}
cvlasso
\optional{,\
\texttt{lopt}\  
\texttt{lse}\
\texttt{plotcv({\method})}\ 
\texttt{plotopt({\it string})}\
\texttt{\underbar{postr}esults}%
}
\end{stsyntax}

\subsubsection{Predict syntax}
\begin{stsyntax}
predict 
\opttype\
{\it newvar}\ 
\optif\ 
\optin\ 
\optional{,\
\texttt{xb}\  
\texttt{\underbar{r}esiduals}\
\texttt{lopt}\ 
\texttt{lse}\ 
\texttt{\underbar{noi}sily}%
}
\end{stsyntax}

\subsection{rlasso: Rigorous penalization}
\rlasso implements theory-driven penalization for lasso and square-root lasso. It allows for heteroskedastic, cluster-dependent and non-Gaussian errors. Unlike \lassotwo and \cvlasso, \rlasso estimates the penalty level $\lambda$ using iterative algorithms. 
The syntax of \rlasso is given by:

\begin{stsyntax}
rlasso 
\depvar\
\textit{indepvars} 
\optif\ 
\optin\
\optweight\
\optional{, 
sqrt  
\underbar{par}tial(\varlist)
\underbar{pnotp}en(\varlist) 
\underbar{nocons}tant 
fe  
\underbar{rob}ust 
\underbar{cl}uster(\var) 
center  
\underbar{xdep}endent 
numsim(\stint)  
prestd 
\underbar{tolo}pt(\real) 
\underbar{tolu}ps(\real)
\underbar{tolz}ero(\real) 
\underbar{maxi}ter(\stint) 
\underbar{maxpsii}ter(\stint) 
lassopsi     
\underbar{corrn}umber(\stint)
maxabsx
\underbar{lalt}ernative
gamma(\real)   
c(\real)  
supscore 
ssnumsim(\stint) 
ssgamma(\real)
testonly 
seed(\stint) 
ols
}
\end{stsyntax}

\par\hangpara
\texttt{\underbar{pnotp}en(\varlist)} specifies that variables in {\it varlist} are not penalized.%
\footnote{This option differs from that of \texttt{\underbar{notp}en(\varlist)} as used with \texttt{cvlasso} and \texttt{lasso2}; see the discussion in Section 9.}
\par\hangpara
\texttt{\underbar{rob}ust}  specifies that the penalty loadings account for heteroskedasticity.
\par\hangpara
\texttt{\underbar{cl}uster({\it varname})} specifies that the penalty loadings account for clustering on variable {\it varname}.
\par\hangpara
\texttt{center} center moments in heteroskedastic and cluster-robust loadings.%
\footnote{For example, the uncentered heteroskedastic loading for regressor $j$ is $\hat\psi_j = \sqrt{\frac{1}{n} \sum_i x_{ij}^2 \hat \varepsilon_i^2}$.
In theory, $x_{ij}\varepsilon_i$ should be mean-zero.
The centered penalty loading is $\hat\psi_j = \sqrt{\frac{1}{n} \sum_i (x_{ij} \hat \varepsilon_i-\hat\mu)^2}$ where $\hat\mu = \frac{1}{n} \sum_i x_{ij}\hat \varepsilon_i$.}
\par\hangpara
\texttt{lassopsi}               use lasso or square-root lasso residuals to obtain penalty loadings. The default is post-estimation OLS.\footnote{The option was called \texttt{lassoups} in earlier versions.}
\par\hangpara
\texttt{\underbar{corrn}umber(\stint)}        number of high-correlation regressors used to obtain initial residuals. The default is \texttt{\underbar{corrn}umber(5)}, and \texttt{\underbar{corrn}umber(0)} implies that {\it depvar} is used in place of residuals.
\par\hangpara
\texttt{prestd} standardize data prior to estimation. The default is  standardization during estimation via penalty loadings.
                            
\subsubsection{Options relating to lambda}
\par\hangpara
\texttt{\underbar{xdep}endent}          specifies that the \emph{X-dependent} penalty level is used; see Section~\ref{sec:rigorous_xdep}.
\par\hangpara
\texttt{numsim(\stint)} is the number of simulations used for the \emph{X-dependent} case. The default is 5,000.
\par\hangpara
\texttt{\underbar{lalt}ernative} specifies the alternative, less sharp penalty level, which is defined as $2c\sqrt{2n\log(2p/\gamma)}$ (for the square-root lasso, $2c$ is replaced with $c$).
See Footnote~\ref{fn:lalt}.
\par\hangpara
\texttt{gamma(\real)} is the `$\gamma$' in the rigorous penalty level (default ${\gamma}=1/{\log(n)}$; cluster-lasso default ${\gamma}=1/{\log(n_{clust})}$).
See Equation~\eqref{eq:idealpenalty}.
\par\hangpara
\texttt{c(\real)} is the `$c$' in the rigorous penalty level.
The default is \texttt{c(1.1)}.
See Equation~\eqref{eq:idealpenalty}.

\subsubsection{Sup-score test}
\par\hangpara
\texttt{supscore} reports the sup-score test of statistical significance.
\par\hangpara
\texttt{testonly} reports only the sup-score test without lasso estimation.
\par\hangpara
\texttt{ssgamma(\real)} is the test level for the conservative critical value for the sup-score test (default = 0.05, i.e., 5\% significance level).
\par\hangpara
\texttt{ssnumsim(\stint)} controls the number of simulations for sup-score test multiplier bootstrap.
The default is 500, while 0 implies no simulation. 

\subsubsection{Predict syntax}
\begin{stsyntax}
predict 
\opttype\
{\it newvar}\ 
\optif\ 
\optin\ 
\optional{,
\texttt{xb}  
\texttt{residuals} 
\texttt{lasso} 
\texttt{ols}%
}
\end{stsyntax}
\par\hangpara
\texttt{xb}                     generate fitted values (default).
\par\hangpara
\texttt{\underbar{r}esiduals}             generate residuals.
\par\hangpara
\texttt{lasso}                  use lasso coefficients for prediction (default is to use estimates posted in \texttt{e(b)} matrix).
\par\hangpara
\texttt{ols}                    use OLS coefficients based on lasso-selected variables for prediction (default is to use estimates posted in \texttt{e(b)} matrix).

\section{Demonstrations}\label{sec:demo}
In this section, we demonstrate the use of \lassotwo, \cvlasso and \rlasso using one cross-section example (in Section~\ref{sec:demo_cs}) and one time-series example (in Section~\ref{sec:demo_ts}). 

\subsection{Cross-section}\label{sec:demo_cs}
For demonstration purposes, we consider the Boston Housing Dataset available on the UCI Machine Learning Repository.\footnote{The dataset is available at \href{https://archive.ics.uci.edu/ml/machine-learning-databases/housing/housing.data}{https://archive.ics.uci.edu/ml/machine-learning-databases/housing/ housing.data}, or in CSV format via our website at \href{http://statalasso.github.io/dta/housing.csv}{http://statalasso.github.io/dta/housing.csv}.}
The data set includes 506 observations and 14 predictors.%
\footnote{The following predictors are included: per capita crime rate (\texttt{crim}), proportion of residential land zoned for lots over 25,000 sq.ft. (\texttt{zn}), proportion of non-retail business acres per town (\texttt{indus}), Charles River dummy variable (\texttt{chas}), nitric oxides concentration (parts per 10 million) (\texttt{nox}), average number of rooms per dwelling (\texttt{rm}), proportion of owner-occupied units built prior to 1940 (\texttt{age}), weighted distances to five Boston employment centres (\texttt{dis}), index of accessibility to radial highways (\texttt{rad}), full-value property-tax rate per \$10,000 (\texttt{tax}), pupil-teacher ratio by town (\texttt{pratio}), $1000(Bk - 0.63)^2$ where Bk is the proportion of blacks by town (\texttt{b}), \% lower status of the population (\texttt{lstat}), median value (\texttt{medv}).}
The purpose of the analysis is to predict house prices using a set of census-level characteristics.

\subsubsection{Estimation with lasso2}
We first employ the lasso estimator: 

\begin{samepage}
\begin{stlog}
. lasso2 medv crim-lstat
{\smallskip}
  Knot{\VBAR}  ID     Lambda    s      L1-Norm        EBIC     R-sq   {\VBAR} Entered/removed
\HLI{6}{\PLUS}\HLI{57}{\PLUS}\HLI{16}
     1{\VBAR}   1 6858.98553     1     0.00000   2250.74087   0.0000  {\VBAR} Added _cons.
     2{\VBAR}   2 6249.65216     2     0.08440   2207.91748   0.0924  {\VBAR} Added lstat.
     3{\VBAR}   3 5694.45029     3     0.28098   2166.62026   0.1737  {\VBAR} Added rm.
     4{\VBAR}  10 2969.09110     4     2.90443   1902.66627   0.5156  {\VBAR} Added ptratio.
     5{\VBAR}  20 1171.07071     5     4.79923   1738.09475   0.6544  {\VBAR} Added b.
     6{\VBAR}  22  972.24348     6     5.15524   1727.95402   0.6654  {\VBAR} Added chas.
     7{\VBAR}  26  670.12972     7     6.46233   1709.14648   0.6815  {\VBAR} Added crim.
     8{\VBAR}  28  556.35346     8     6.94988   1705.73465   0.6875  {\VBAR} Added dis.
     9{\VBAR}  30  461.89442     9     8.10548   1698.65787   0.6956  {\VBAR} Added nox.
    10{\VBAR}  34  318.36591    10    13.72934   1679.28783   0.7106  {\VBAR} Added zn.
    11{\VBAR}  39  199.94307    12    18.33494   1671.61672   0.7219  {\VBAR} Added indus rad.
    12{\VBAR}  41  165.99625    13    20.10743   1669.76857   0.7263  {\VBAR} Added tax.
    13{\VBAR}  47   94.98916    12    23.30144   1645.44345   0.7359  {\VBAR} Removed indus.
    14{\VBAR}  67   14.77724    13    26.71618   1642.91756   0.7405  {\VBAR} Added indus.
    15{\VBAR}  82    3.66043    14    27.44510   1648.83626   0.7406  {\VBAR} Added age.
Use 'long' option for full output. 
Type e.g. '\textcolor{blue}{lasso2, lic(ebic)}' to run the model selected by EBIC.
\end{stlog}
\end{samepage}

\noindent The above \texttt{lasso2} output shows the following columns: 
\begin{itemize}[nosep] 
    \item \texttt{Knot} is the knot index. Knots are points at which predictors enter or leave the model. The default output shows one line per knot. If the \texttt{long} option is specified, one row per $\lambda_r$ value is shown. 
    \item \texttt{ID} shows the $\lambda_r$ index, i.e., $r$. By default, \texttt{lasso2} uses a descending sequence of 100 penalty levels.
    \item \texttt{s} is the number of predictors in the model. 
    \item \texttt{L1-Norm} shows the $\ell_1$-norm of coefficient estimates. 
    \item The sixth column (here labelled \texttt{EBIC}) shows one out of four information criteria. The \myoption{ic}{\ststring} option controls which information criterion is displayed, where \ststring{} can be replaced with `aic', `aicc', `bic', and `ebic' (the default).
    \item \texttt{R-sq} shows the $R^2$ value.
    \item The final column shows which predictors are entered or removed from the model at each knot. The order in which predictors are entered into the model can be interpreted as an indication of the relative predictive power of each predictor. 
\end{itemize}

\begin{figure}
    \centering
    \begin{minipage}{.495\linewidth}
    \sjversion{\includegraphics[width=\linewidth]{plotpath_housing_mono.pdf}}{\includegraphics[width=\linewidth]{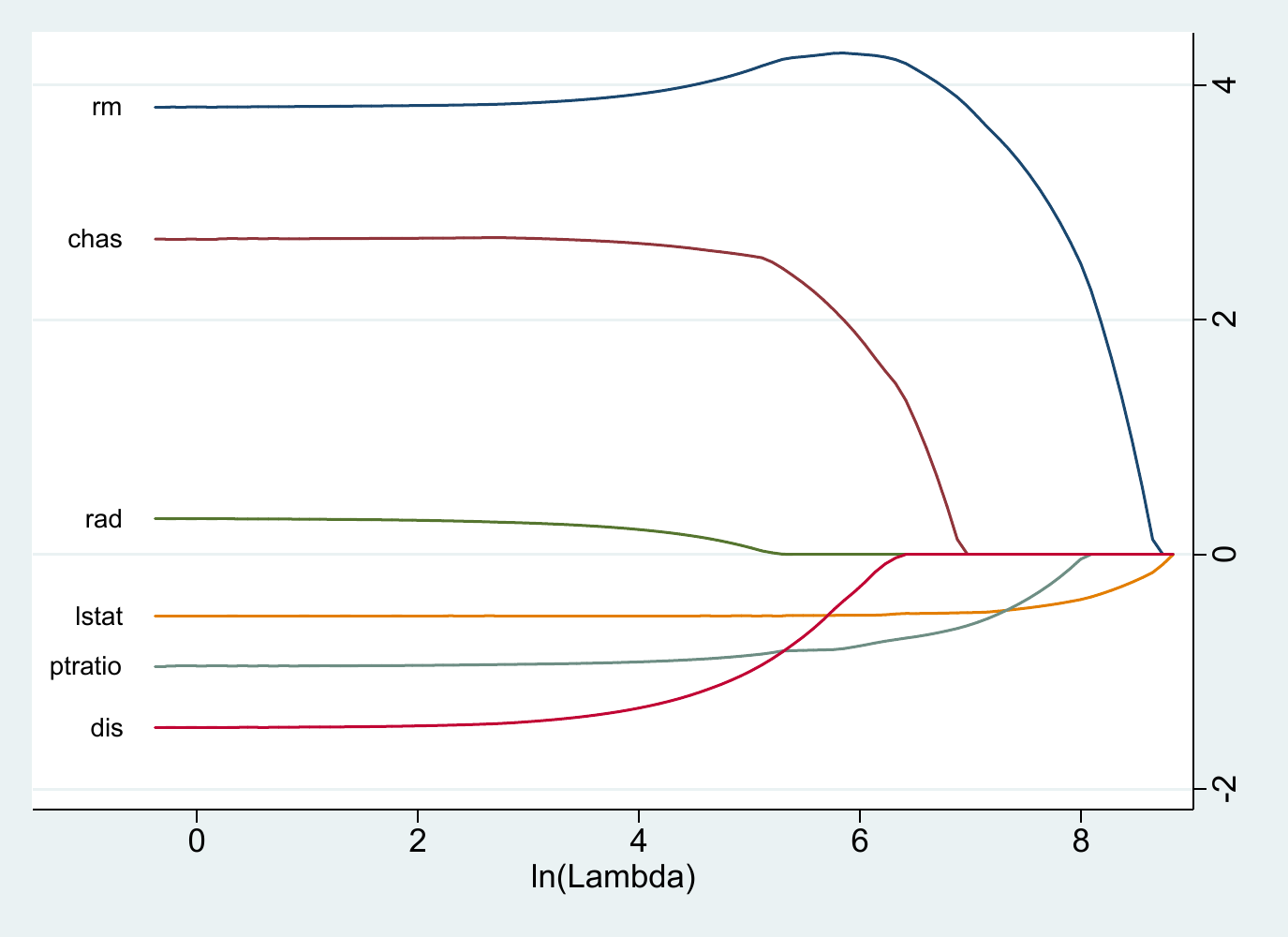}}
    \subcaption{Coefficient path\label{fig:coef_path}}
    \end{minipage}\hfill%
    \begin{minipage}{.495\linewidth}
    \sjversion{\includegraphics[width=\linewidth]{plotcv_housing_mono.pdf}}{\includegraphics[width=\linewidth]{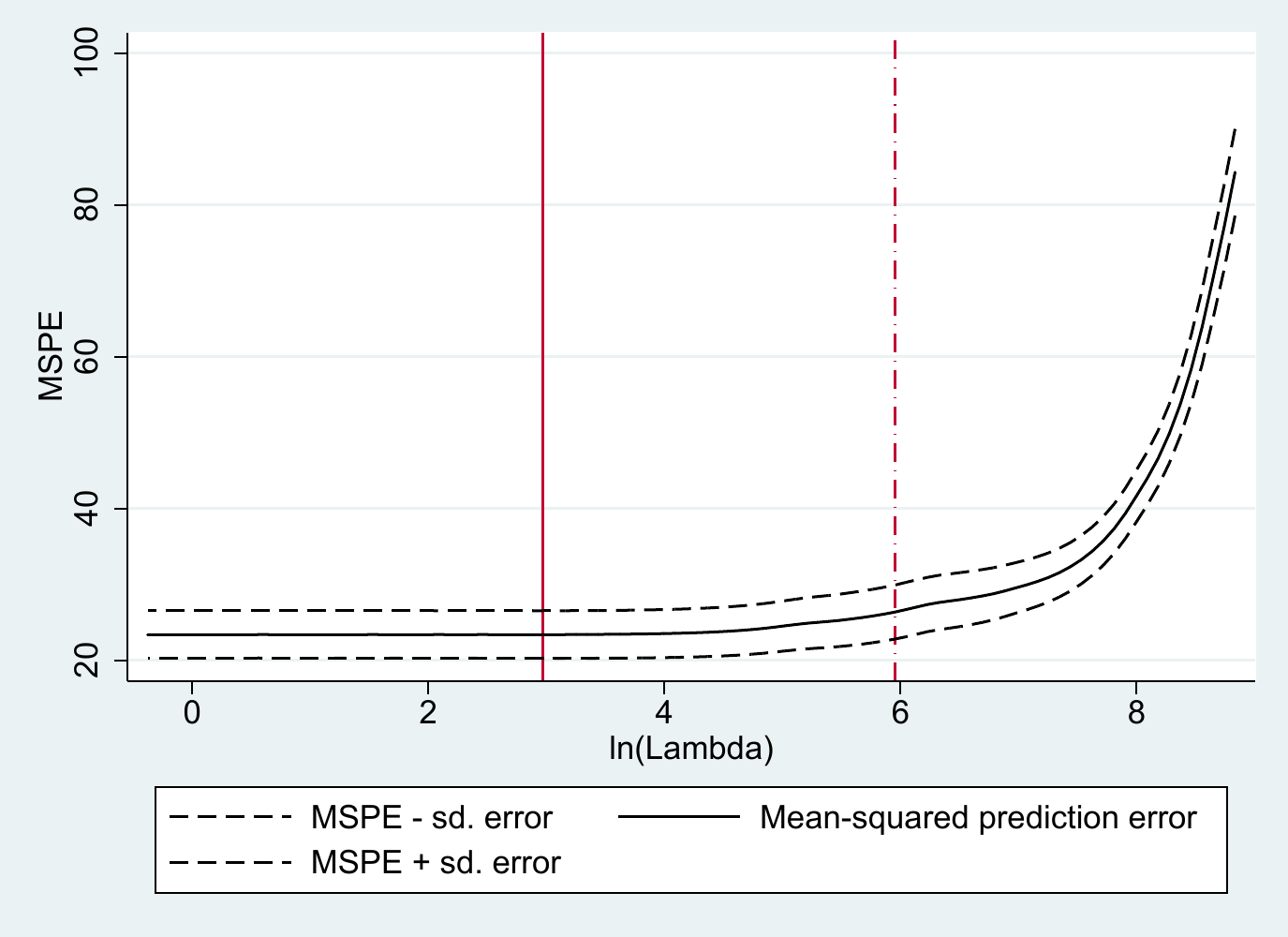}}
    \subcaption{Mean squared prediction error}\label{fig:cv}
    \end{minipage}
    \caption{The left graph shows the coefficient path of the lasso for selected variables as a function of $\ln(\lambda)$. The right graph shows the mean squared prediction error estimated by cross-validation along with $\pm$ one standard error. The continuous and dashed vertical lines correspond to \texttt{lopt} and \texttt{lse}, respectively.}
\end{figure}

Since \texttt{lambda(\real)} is not specified, \lassotwo obtains the coefficient path for a default list of $\lambda_r$ values. The largest penalty level is 6858.99, in which case the model does only include the constant. Figure~\ref{fig:coef_path}  shows the coefficient path of the lasso for selected variables as a function of $\ln(\lambda)$.\footnote{Figure~\ref{fig:coef_path} was created using the following command:\\
\texttt{.\ lasso2 medv crim-lstat, plotpath(lnlambda) plotopt(legend(off)) plotlabel plotvar(rm chas rad lstat ptratio dis)}}

\texttt{lasso2} supports model selection using information criteria. To this end, we use the replay syntax in combination with the \texttt{lic()} option, which avoids that the full model needs to be estimated again. The \texttt{lic()} option can also be specified in the first \lassotwo call. In the following example, the replay syntax works similar to a post-estimation command. 

\begin{samepage}
\begin{stlog}
. lasso2, lic(ebic)
{\smallskip}
Use lambda=16.21799867742649 (selected by EBIC).
{\smallskip}
\HLI{18}{\TOPT}\HLI{32}
         Selected {\VBAR}           Lasso   Post-est OLS
\HLI{18}{\PLUS}\HLI{32}
             crim {\VBAR}      -0.1028391     -0.1084133
               zn {\VBAR}       0.0433716      0.0458449
             chas {\VBAR}       2.6983218      2.7187164
              nox {\VBAR}     -16.7712529    -17.3760262
               rm {\VBAR}       3.8375779      3.8015786
              dis {\VBAR}      -1.4380341     -1.4927114
              rad {\VBAR}       0.2736598      0.2996085
              tax {\VBAR}      -0.0106973     -0.0117780
          ptratio {\VBAR}      -0.9373015     -0.9465246
                b {\VBAR}       0.0091412      0.0092908
            lstat {\VBAR}      -0.5225124     -0.5225535
\HLI{18}{\PLUS}\HLI{32}
   Partialled-out*{\VBAR}
\HLI{18}{\PLUS}\HLI{32}
            _cons {\VBAR}      35.2705812     36.3411478
\HLI{18}{\BOTT}\HLI{32}
\end{stlog}
\end{samepage}

Two columns are shown in the output; one for the lasso estimator and one for post-estimation OLS, which applies OLS to the model selected by the lasso.

\subsubsection{K-fold cross-validation with cvlasso}
Next, we consider $K$-fold cross-validation. 

\begin{samepage}
\begin{stlog}
. set seed 123
{\smallskip}
. cvlasso medv crim-lstat
{\smallskip}
K-fold cross-validation with 10 folds. Elastic net with alpha=1.
Fold 1 2 3 4 5 6 7 8 9 10 
          {\VBAR}         Lambda           MSPE       st. dev.
\HLI{10}{\PLUS}\HLI{45}
         1{\VBAR}      6858.9855      84.302552      5.7124688  
         2{\VBAR}      6249.6522      77.022038      5.5626292  
         3{\VBAR}      5694.4503      70.352232      5.3037622  
\textnormal{\hspace*{.3cm}\emph{(Output omitted.)}}
        30{\VBAR}      461.89442      27.034557      3.5821586  
        31{\VBAR}      420.86099      26.695961      3.5812873  
        32{\VBAR}      383.47286      26.365176      3.5552884  {\caret}
        33{\VBAR}      349.40619      26.095202      3.5350981  
        34{\VBAR}      318.36591      25.857426        3.51782 
\textnormal{\hspace*{.3cm}\emph{(Output omitted.)}}
        62{\VBAR}      23.529539      23.421433      3.1339813  
        63{\VBAR}       21.43924      23.419627       3.131822  
        64{\VBAR}      19.534637      23.418936      3.1298343  *
        65{\VBAR}      17.799234      23.419177      3.1280902  
        66{\VBAR}      16.217999      23.419668      3.1266572 
\textnormal{\hspace*{.3cm}\emph{(Output omitted.)}}
        98{\VBAR}      .82616724      23.441147      3.1134727  
        99{\VBAR}      .75277282      23.441321      3.1134124  
       100{\VBAR}      .68589855      23.441481      3.1133575  
* lopt = the lambda that minimizes MSPE.
  Run model: \textcolor{blue}{cvlasso, lopt}
{\caret} lse = largest lambda for which MSPE is within one standard error of the minimal MSPE.
  Run model: \textcolor{blue}{cvlasso, lse}
\end{stlog}
\end{samepage}

The \cvlasso output displays four columns: 
 the index of $\lambda_r$ (i.e.,  $r$),
  the value of $\lambda_r$, 
   the estimated mean squared prediction error,
    and the standard deviation of the mean squared prediction error.
The output indicates the value of $\lambda_r$ that corresponds to the lowest MSPE with an asterisk (\verb+*+). We refer to this value as $\hat\lambda_{\texttt{lopt}}$. In addition, the symbol \verb+^+ marks the largest value of $\lambda$ that is within one standard error of $\hat\lambda_{\texttt{lopt}}$, which we denote as $\hat\lambda_{\texttt{lse}}$. 

The mean squared prediction is shown in Figure~\ref{fig:cv}, which was created using the \texttt{plotcv} option. The graph shows the mean squared prediction error estimated by cross-validation along with $\pm$ one standard error. The continuous and dashed vertical lines correspond to $\hat\lambda_{\texttt{lopt}}$ and $\hat\lambda_{\texttt{lse}}$, respectively.

To estimate the model corresponding to either $\hat\lambda_{\texttt{lopt}}$ or  $\hat\lambda_{\texttt{lse}}$, we use the \texttt{lopt} or \texttt{lse} option, respectively. Similar to the \texttt{lic()} option of \lassotwo, \texttt{lopt} and \texttt{lse} can either specified in the first \cvlasso call or after estimation using the replay syntax as in this example:

\vspace*{-.3cm}

\begin{samepage}
\begin{stlog}
. cvlasso, lopt
Estimate lasso with lambda=19.535 (lopt).
{\smallskip}
\HLI{18}{\TOPT}\HLI{32}
         Selected {\VBAR}           Lasso   Post-est OLS
\HLI{18}{\PLUS}\HLI{32}
             crim {\VBAR}      -0.1016991     -0.1084133
               zn {\VBAR}       0.0428658      0.0458449
             chas {\VBAR}       2.6941511      2.7187164
              nox {\VBAR}     -16.6475746    -17.3760262
               rm {\VBAR}       3.8449399      3.8015786
              dis {\VBAR}      -1.4268524     -1.4927114
              rad {\VBAR}       0.2683532      0.2996085
              tax {\VBAR}      -0.0104763     -0.0117780
          ptratio {\VBAR}      -0.9354154     -0.9465246
                b {\VBAR}       0.0091106      0.0092908
            lstat {\VBAR}      -0.5225040     -0.5225535
\HLI{18}{\PLUS}\HLI{32}
   Partialled-out*{\VBAR}
\HLI{18}{\PLUS}\HLI{32}
            _cons {\VBAR}      35.0516465     36.3411478
\HLI{18}{\BOTT}\HLI{32}
\end{stlog}
\end{samepage}

\subsubsection{Rigorous penalization with rlasso}
Lastly, we consider \rlasso. The program \rlasso runs an iterative algorithm to estimate the penalty level and loadings. In contrast to \lassotwo and \cvlasso, it reports the selected model directly.

\begin{samepage}
\begin{stlog}
. rlasso medv crim-lstat, supscore
{\smallskip}
\HLI{18}{\TOPT}\HLI{32}
         Selected {\VBAR}           Lasso   Post-est OLS
\HLI{18}{\PLUS}\HLI{32}
             chas {\VBAR}       0.6614716      3.3200252
               rm {\VBAR}       4.0224498      4.6522735
          ptratio {\VBAR}      -0.6685443     -0.8582707
                b {\VBAR}       0.0036058      0.0101119
            lstat {\VBAR}      -0.5009804     -0.5180622
            _cons {\VBAR}{\bftt{*}}     14.5986089     11.8535884
\HLI{18}{\BOTT}\HLI{32}
*Not penalized
{\smallskip}
Sup-score test H0: beta=0
CCK sup-score statistic  16.59 p-value= 0.000
CCK 5\% critical value     3.18 (asympt bound)
\end{stlog}
\end{samepage}

The \texttt{supscore} option prompts the sup-score test of joint significance. The $p$-value is obtained through multiplier bootstrap. The test statistic of 16.59 can also be compared to the asymptotic 5\% critical value (here 3.18). 

\subsection{Time-series data}\label{sec:demo_ts}
A standard problem in time-series econometrics is to select an appropriate lag length. In this sub-section, we show how \lassopack can be employed for this purpose.
We consider Stata's built-in data set \texttt{lutkepohl2.dta}, which includes quarterly (log-differenced) consumption (\texttt{dln\_consump}), investment (\texttt{dln\_inv}) and income (\texttt{dln\_inc}) series for West Germany over the period 1960, Quarter 1 to 1982, Quarter 4. 
We demonstrate both lag selection via information criteria and by $h$-step ahead rolling cross-validation. We do not consider the rigorous penalization approach of \texttt{rlasso} due to the assumption of independence, which seems too restrictive in the time-series context. 

\subsubsection{Information criteria}
After importing the data, we run the most general model with up to 12 lags of \texttt{dln\_consump}, \texttt{dln\_inv} and \texttt{dln\_inc} using \lassotwo with \texttt{lic(aicc)} option. 

\begin{samepage}
\begin{stlog}
. lasso2 dln_consump L(1/12).(dln_inv dln_inc dln_consump), lic(aicc) long
{\smallskip}
  Knot{\VBAR}  ID     Lambda    s      L1-Norm        AICc     R-sq   {\VBAR} Entered/removed
\HLI{6}{\PLUS}\HLI{57}{\PLUS}\HLI{16}
     1{\VBAR}   1    0.52531     1     0.00000   -714.43561   0.0000  {\VBAR} Added _cons.
\textnormal{\hspace*{.3cm}\emph{(Output omitted.)}}
      {\VBAR}  11    0.20719    10     0.67593   -722.62355*  0.3078  {\VBAR}
\textnormal{\hspace*{.3cm}\emph{(Output omitted.)}}
      {\VBAR} 100    0.00005    37     4.92856   -665.31816   0.6719  {\VBAR}
{\bftt{*}}indicates minimum AICc.
Use lambda=.2071920751852477 (selected by AICC).
{\smallskip}
\HLI{18}{\TOPT}\HLI{32}
         Selected {\VBAR}           Lasso   Post-est OLS
\HLI{18}{\PLUS}\HLI{32}
          dln_inv {\VBAR}
              L2. {\VBAR}       0.0279780      0.0513004
                  {\VBAR}
          dln_inc {\VBAR}
              L1. {\VBAR}       0.0672531      0.1522251
              L2. {\VBAR}       0.1184912      0.1675746
              L3. {\VBAR}       0.0779780      0.1261940
              L8. {\VBAR}      -0.1091959     -0.2481821
                  {\VBAR}
      dln_consump {\VBAR}
              L2. {\VBAR}       0.0259311      0.0935048
              L3. {\VBAR}       0.0765755      0.1405377
             L10. {\VBAR}       0.0833425      0.2320500
             L11. {\VBAR}      -0.0891871     -0.1442602
\HLI{18}{\PLUS}\HLI{32}
   Partialled-out*{\VBAR}
\HLI{18}{\PLUS}\HLI{32}
                  {\VBAR}
            _cons {\VBAR}       0.0133270      0.0079518
\HLI{18}{\BOTT}\HLI{32}
\end{stlog}
\end{samepage}

The output consists of two parts.  The second part of the output is prompted since \stcmd{lic(aicc)} is specified. \stcmd{lic(aicc)} asks \lassotwo to estimate the model selected by AIC$_c$, which in this case corresponds to $\lambda_{11}=0.207$. 

\subsubsection{h-step ahead rolling cross-validation}
In the next step, we consider $h$-step ahead rolling cross-validation. 

\begin{figure}
    \centering
    \sjversion{\includegraphics[width=.6\linewidth]{plotcv_lutkepohl_mono.pdf}}{\includegraphics[width=.6\linewidth]{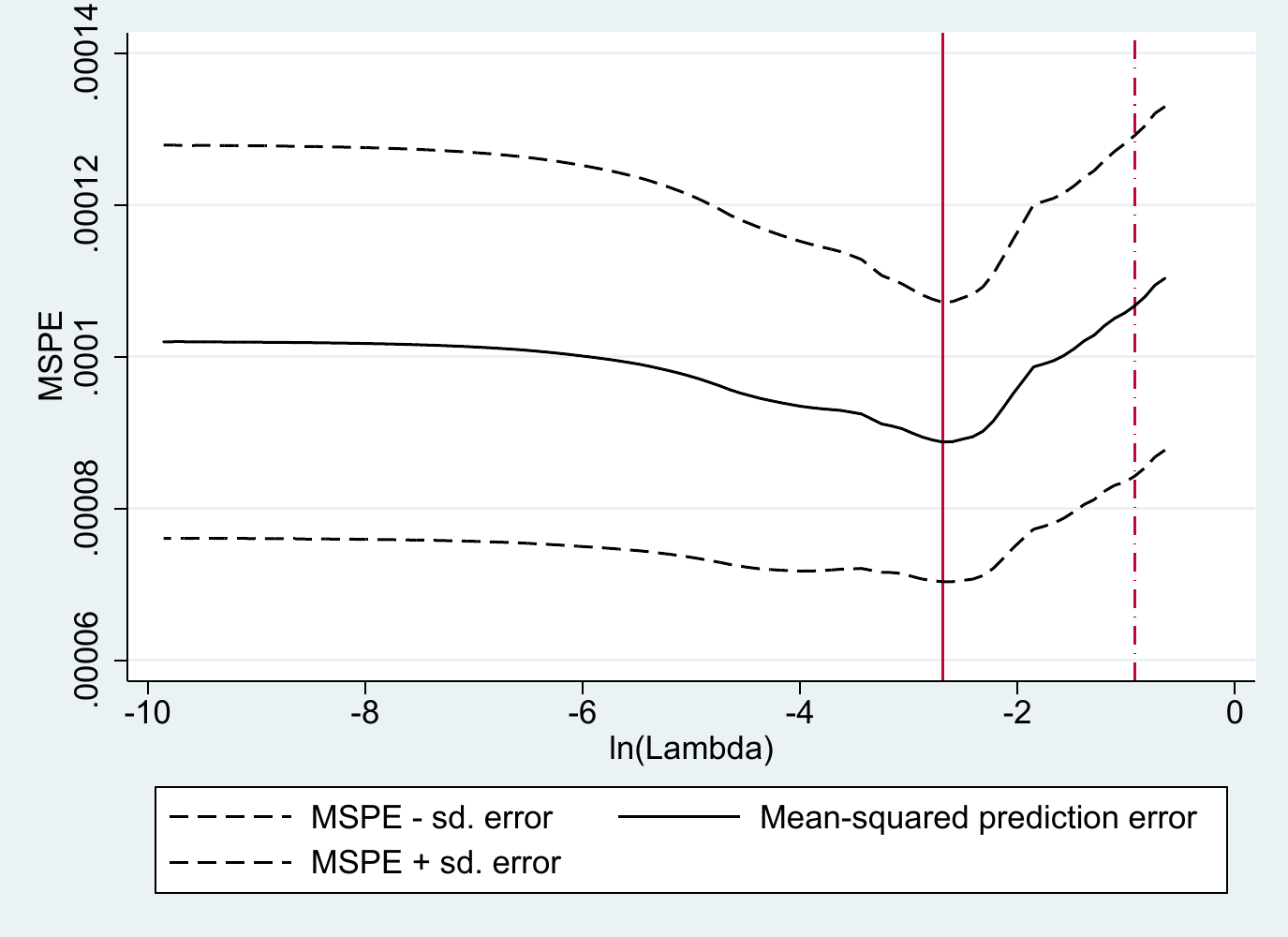}}
    \caption{Cross-validation plot. The graph uses Stata's built-in data set \texttt{lutkepohl2.dta} and 1-step ahead rolling cross-validation with \stcmd{origin(50)}.}\label{fig:cvlasso_timeseries}
\end{figure}

\begin{stlog}
. cvlasso dln_consump L(1/12).(dln_inv dln_inc dln_consump), rolling
Rolling forecasting cross-validation with 1-step ahead forecasts. Elastic net with alpha=1.
Training from-to (validation point): 13-80 (81), 13-81 (82), 13-82 (83), 13-83 (84),
13-84 (85), 13-85 (86), 13-86 (87), 13-87 (88), 13-88 (89), 13-89 (> 90), 13-90 (91).
\end{stlog}

The output indicates how the data set is partitioned into training data and the validation point. For example, the short-hand \texttt{13-80 (81)} in the output above indicates that observations 13 to 80 constitute the training data set in the first step of cross-validation, and observation 81 is the validation point. The options \stcmd{fixedwindow}, \myoption{h}{\textit{integer}} and \myoption{origin}{\textit{integer}} can be used to control the partitioning of data into training and validation data. \myoption{h}{\textit{integer}} sets the parameter $h$. For example, \stcmd{h(2)} prompts 2-step ahead forecasts (the default is \stcmd{h(1)}). \stcmd{fixedwindow} ensures that the training data set is of same size in each step. If \stcmd{origin(50)} is specified, the first training partition includes observations 13 to 50, as shown in the next example:

\begin{stlog}
. cvlasso dln_consump L(1/12).(dln_inv dln_inc dln_consump), rolling origin(50) plotcv
Rolling forecasting cross-validation with 1-step ahead forecasts. Elastic net with alpha=1.
Training from-to (validation point): 13-50 (51), 13-51 (52), 13-52 (53), 13-53 (54),
\textnormal{\hspace*{.3cm}\emph{(Output omitted.)}}
13-82 (83), 13-83 (84), 13-84 (85), 13-85 (86), 13-86 (87), 13-87 (88), 13-88 (89),
13-89 (90), 13-90 (91).
\end{stlog}

The option \texttt{plotcv} creates the graph of the estimated mean squared prediction in Figure~\ref{fig:cvlasso_timeseries}. To estimate the model corresponding to $\hat\lambda_{\texttt{lse}}$, we can as in the previous examples use the replay syntax:

\begin{samepage}
\begin{stlog}
. cvlasso, lse
Estimate lasso with lambda=.397 (lse).
{\smallskip}
\HLI{18}{\TOPT}\HLI{32}
         Selected {\VBAR}           Lasso   Post-est OLS
\HLI{18}{\PLUS}\HLI{32}
          dln_inv {\VBAR}
              L2. {\VBAR}       0.0071068      0.0481328
                  {\VBAR}
          dln_inc {\VBAR}
              L2. {\VBAR}       0.0558422      0.2083321
              L3. {\VBAR}       0.0253076      0.1479925
                  {\VBAR}
      dln_consump {\VBAR}
              L3. {\VBAR}       0.0260573      0.1079076
             L11. {\VBAR}      -0.0299307     -0.1957719
\HLI{18}{\PLUS}\HLI{32}
   Partialled-out*{\VBAR}
\HLI{18}{\PLUS}\HLI{32}
                  {\VBAR}
            _cons {\VBAR}       0.0168736      0.0126810
\HLI{18}{\BOTT}\HLI{32}
\end{stlog}
\end{samepage}

We point out that care should be taken when setting the parameters of $h$-step ahead rolling cross-validation. The default settings have no particular econometric justification.

\section{Monte Carlo Simulation}\label{sec:mc}
We have introduced three alternative approaches for setting the penalization parameters in Sections~\ref{sec:informationcriteria}-\ref{sec:rigorous}. In this section, we present results of Monte Carlo simulations which assess the performance of these approaches in terms of in-sample fit, out-of-sample prediction, model selection and sparsity. To this end, we generate artificial data using the process
\begin{align}
y_i=1+\sum_{j=1}^p \beta_j x_{ij} + \varepsilon_i,  \quad \varepsilon_i\sim \mathcal{N}(0,\sigma^2), \quad i=1,\ldots, 2\obs, 
\end{align}
with $\obs=200$. We report results for $p=100$ and for the high-dimensional setting with $p=220$.  The predictors $x_{ij}$ are drawn from a multivariate normal distribution with $\textnormal{corr}(x_{ij},x_{ir})=0.9^{|j-r|}$. We vary the noise level $\sigma$ between $0.5$ and $5$; specifically, we consider $\sigma=\{0.5,1,2,3,5\}$.
We define the parameters as $\beta_j = \mathbbm{1}\{j\leq s\}$ for $j=1,\ldots,p$ with $s=20$, implying exact sparsity. This simple design allows us to gain insights into the model selection performance in terms of false positive (the number of variables falsely included) and false negative frequency (the number of variables falsely omitted) when relevant and irrelevant regressors are correlated. All simulations use at least 1,000 iterations. We report additional Monte Carlo results in Appendix~\ref{sec:moremc}, where we employ a design in which coefficients alternate in sign.

Since the aim is to assess in-sample and out-of-sample performance, we generate $2\obs$ observations, and use the data $i=1,\ldots,\obs$ as the estimation sample and the observations $i=\obs+1,\ldots,2\obs$ for assessing out-of-sample prediction performance. This allows us to calculate the root mean squared error (RMSE) and root mean squared prediction error (RMSPE) as 
\begin{equation}
 \textnormal{RMSE} = \sqrt{\frac{1}{\obs}\sum^\obs_{i=1} (y_i - \hat{y}_{i,\obs})^2} \qquad\textnormal{and}\qquad 
 \textnormal{RMSPE} = \sqrt{\frac{1}{\obs}\sum^{2\obs}_{i=\obs+1} (y_i - \hat{y}_{i,\obs})^2}, \label{eq:rmse_rmspe}
\end{equation}
where $\hat{y}_{i,\obs}$ are the predictions from fitting the model to the first $\obs$ observations.

Table~\ref{tab:mc_design1} and~\ref{tab:mc_design1_hd} report results for the following estimation methods implemented in \lassopack: the lasso with $\lambda$ selected by AIC, BIC, EBIC$_\xi$ and AIC$_c$ (as implemented in \lassotwo); the rigorous lasso and rigorous square-root lasso (implemented in \rlasso), both using the \emph{X-independent} and \emph{X-dependent} penalty choice; and lasso with 5-fold cross-validation using the penalty level that minimizes the estimated mean squared prediction error (implemented in \cvlasso). In addition, we report post-estimation OLS results. 

\begin{table}
    \centering\singlespacing
    \newcolumntype{C}{>{\centering\arraybackslash}X}
    \newcommand{\myrotate}[1]{\multirow{5}{*}{\rotatebox{90}{#1}\hspace*{-3pt}}}
        \newcommand{\specialcell}[2][t]{\begin{tabular}[#1]{@{}c@{}}#2\end{tabular}}
    \scriptsize
    \begin{tabularx}{\textwidth}{llCCCCC@{\hspace{1.9\tabcolsep}}CCCCCCCC}
    \toprule \midrule
     &$\sigma$&\multicolumn{4}{l}{\lassotwo}&\multicolumn{1}{l}{\cvlasso}&\multicolumn{4}{l}{\rlasso}&Step&{Oracle} \tabularnewline
     & &{AIC}&{AIC$_c$}&{BIC}&{EBIC$_\xi$}&&\multicolumn{2}{l}{lasso} &\multicolumn{2}{l}{$\sqrt{\textnormal{lasso}}$}&wise&\tabularnewline
      & & &  & & & &  & \texttt{xdep} &   & \texttt{xdep} &\tabularnewline
    \midrule\addlinespace[1.5ex]
    \myrotate{$\hat{s}$}&.5&38.14&24.57&21.47&20.75&25.54&20.19&20.22&20.23&20.27&37.26&-- \tabularnewline
&1&38.62&24.50&21.50&20.73&25.56&20.27&20.30&20.25&20.27&37.23&-- \tabularnewline
&2&38.22&24.37&20.98&20.26&25.51&19.78&19.83&19.68&19.74&33.27&-- \tabularnewline
&3&36.94&23.17&19.69&18.83&24.13&18.32&18.39&18.05&18.17&30.15&-- \tabularnewline
&5&33.35&20.46&16.52&15.52&21.23&15.13&15.25&14.70&14.86&27.90&-- \tabularnewline
\midrule \myrotate{False pos.}&.5&18.14&4.57&1.47&0.75&5.54&0.19&0.22&0.23&0.27&18.26&-- \tabularnewline
&1&18.62&4.50&1.50&0.73&5.56&0.28&0.30&0.25&0.28&18.53&-- \tabularnewline
&2&18.64&4.75&1.38&0.71&5.86&0.28&0.32&0.22&0.25&20.17&-- \tabularnewline
&3&18.73&4.86&1.48&0.75&5.71&0.32&0.35&0.26&0.29&20.14&-- \tabularnewline
&5&17.87&5.03&1.29&0.58&5.58&0.25&0.28&0.19&0.22&20.58&-- \tabularnewline
\midrule \myrotate{False neg.}&.5&0.00&0.00&0.00&0.00&0.00&0.00&0.00&0.00&0.00&0.00&-- \tabularnewline
&1&0.00&0.00&0.00&0.00&0.00&0.00&0.00&0.00&0.00&0.30&-- \tabularnewline
&2&0.42&0.38&0.41&0.45&0.35&0.50&0.48&0.55&0.52&5.90&-- \tabularnewline
&3&1.78&1.68&1.79&1.92&1.58&2.00&1.96&2.20&2.12&8.99&-- \tabularnewline
&5&4.52&4.57&4.77&5.06&4.35&5.12&5.03&5.50&5.36&11.67&-- \tabularnewline
\midrule \myrotate{Bias}&.5&\specialcell{3.139\\ (\emph{4.155})}&\specialcell{2.002\\ (\emph{2.208})}&\specialcell{1.910\\ (\emph{1.952})}&\specialcell{1.898\\ (\emph{1.894})}&\specialcell{1.999\\ (\emph{2.265})}&\specialcell{2.074\\ (\emph{1.835})}&\specialcell{2.041\\ (\emph{1.838})}&\specialcell{1.990\\ (\emph{1.842})}&\specialcell{1.972\\ (\emph{1.846})}&5.529\newline (--)&1.803\newline (--) \tabularnewline
&1&\specialcell{6.421\\ (\emph{8.459})}&\specialcell{3.974\\ (\emph{4.374})}&\specialcell{3.798\\ (\emph{3.891})}&\specialcell{3.771\\ (\emph{3.756})}&\specialcell{3.983\\ (\emph{4.522})}&\specialcell{3.964\\ (\emph{3.670})}&\specialcell{3.931\\ (\emph{3.674})}&\specialcell{3.958\\ (\emph{3.663})}&\specialcell{3.922\\ (\emph{3.667})}&11.379\newline (--)&3.578\newline (--) \tabularnewline
&2&\specialcell{12.510\\ (\emph{16.563})}&\specialcell{7.836\\ (\emph{8.741})}&\specialcell{7.412\\ (\emph{7.601})}&\specialcell{7.370\\ (\emph{7.402})}&\specialcell{7.847\\ (\emph{9.015})}&\specialcell{7.607\\ (\emph{7.277})}&\specialcell{7.569\\ (\emph{7.281})}&\specialcell{7.670\\ (\emph{7.283})}&\specialcell{7.609\\ (\emph{7.272})}&28.241\newline (--)&7.117\newline (--) \tabularnewline
&3&\specialcell{18.294\\ (\emph{24.274})}&\specialcell{11.093\\ (\emph{12.427})}&\specialcell{10.461\\ (\emph{10.913})}&\specialcell{10.377\\ (\emph{10.641})}&\specialcell{11.134\\ (\emph{12.828})}&\specialcell{10.501\\ (\emph{10.542})}&\specialcell{10.470\\ (\emph{10.530})}&\specialcell{10.570\\ (\emph{10.633})}&\specialcell{10.518\\ (\emph{10.572})}&41.475\newline (--)&10.679\newline (--) \tabularnewline
&5&\specialcell{26.965\\ (\emph{36.638})}&\specialcell{15.878\\ (\emph{18.423})}&\specialcell{14.529\\ (\emph{15.558})}&\specialcell{14.371\\ (\emph{15.326})}&\specialcell{15.666\\ (\emph{18.617})}&\specialcell{14.292\\ (\emph{15.337})}&\specialcell{14.277\\ (\emph{15.271})}&\specialcell{14.343\\ (\emph{15.582})}&\specialcell{14.313\\ (\emph{15.442})}&63.995\newline (--)&18.040\newline (--) \tabularnewline
\midrule \myrotate{RMSE}&.5&\specialcell{0.433\\ (\emph{0.421})}&\specialcell{0.466\\ (\emph{0.456})}&\specialcell{0.479\\ (\emph{0.467})}&\specialcell{0.484\\ (\emph{0.470})}&\specialcell{0.466\\ (\emph{0.454})}&\specialcell{0.546\\ (\emph{0.473})}&\specialcell{0.536\\ (\emph{0.473})}&\specialcell{0.522\\ (\emph{0.473})}&\specialcell{0.517\\ (\emph{0.473})}&0.403\newline (--)&0.474\newline (--) \tabularnewline
&1&\specialcell{0.862\\ (\emph{0.838})}&\specialcell{0.930\\ (\emph{0.909})}&\specialcell{0.955\\ (\emph{0.931})}&\specialcell{0.967\\ (\emph{0.938})}&\specialcell{0.931\\ (\emph{0.905})}&\specialcell{1.041\\ (\emph{0.943})}&\specialcell{1.031\\ (\emph{0.943})}&\specialcell{1.042\\ (\emph{0.943})}&\specialcell{1.030\\ (\emph{0.943})}&0.803\newline (--)&0.945\newline (--) \tabularnewline
&2&\specialcell{1.724\\ (\emph{1.675})}&\specialcell{1.856\\ (\emph{1.815})}&\specialcell{1.912\\ (\emph{1.863})}&\specialcell{1.935\\ (\emph{1.877})}&\specialcell{1.859\\ (\emph{1.807})}&\specialcell{2.057\\ (\emph{1.887})}&\specialcell{2.042\\ (\emph{1.887})}&\specialcell{2.082\\ (\emph{1.888})}&\specialcell{2.059\\ (\emph{1.888})}&1.620\newline (--)&1.891\newline (--) \tabularnewline
&3&\specialcell{2.589\\ (\emph{2.518})}&\specialcell{2.785\\ (\emph{2.725})}&\specialcell{2.871\\ (\emph{2.796})}&\specialcell{2.914\\ (\emph{2.818})}&\specialcell{2.791\\ (\emph{2.716})}&\specialcell{3.080\\ (\emph{2.833})}&\specialcell{3.059\\ (\emph{2.832})}&\specialcell{3.123\\ (\emph{2.836})}&\specialcell{3.089\\ (\emph{2.834})}&2.437\newline (--)&2.836\newline (--) \tabularnewline
&5&\specialcell{4.356\\ (\emph{4.236})}&\specialcell{4.659\\ (\emph{4.554})}&\specialcell{4.819\\ (\emph{4.690})}&\specialcell{4.904\\ (\emph{4.727})}&\specialcell{4.678\\ (\emph{4.551})}&\specialcell{5.146\\ (\emph{4.749})}&\specialcell{5.113\\ (\emph{4.747})}&\specialcell{5.220\\ (\emph{4.756})}&\specialcell{5.165\\ (\emph{4.752})}&4.054\newline (--)&4.730\newline (--) \tabularnewline
\midrule \myrotate{RMSPE}&.5&\specialcell{0.558\\ (\emph{0.589})}&\specialcell{0.539\\ (\emph{0.548})}&\specialcell{0.540\\ (\emph{0.536})}&\specialcell{0.543\\ (\emph{0.533})}&\specialcell{0.539\\ (\emph{0.550})}&\specialcell{0.605\\ (\emph{0.529})}&\specialcell{0.594\\ (\emph{0.529})}&\specialcell{0.580\\ (\emph{0.529})}&\specialcell{0.574\\ (\emph{0.529})}&0.623\newline (--)&0.528\newline (--) \tabularnewline
&1&\specialcell{1.120\\ (\emph{1.181})}&\specialcell{1.078\\ (\emph{1.096})}&\specialcell{1.081\\ (\emph{1.073})}&\specialcell{1.087\\ (\emph{1.065})}&\specialcell{1.078\\ (\emph{1.100})}&\specialcell{1.158\\ (\emph{1.060})}&\specialcell{1.148\\ (\emph{1.060})}&\specialcell{1.159\\ (\emph{1.059})}&\specialcell{1.147\\ (\emph{1.060})}&1.259\newline (--)&1.057\newline (--) \tabularnewline
&2&\specialcell{2.231\\ (\emph{2.355})}&\specialcell{2.149\\ (\emph{2.189})}&\specialcell{2.155\\ (\emph{2.139})}&\specialcell{2.168\\ (\emph{2.125})}&\specialcell{2.149\\ (\emph{2.199})}&\specialcell{2.280\\ (\emph{2.115})}&\specialcell{2.265\\ (\emph{2.115})}&\specialcell{2.305\\ (\emph{2.115})}&\specialcell{2.282\\ (\emph{2.114})}&2.621\newline (--)&2.110\newline (--) \tabularnewline
&3&\specialcell{3.325\\ (\emph{3.509})}&\specialcell{3.201\\ (\emph{3.263})}&\specialcell{3.211\\ (\emph{3.191})}&\specialcell{3.235\\ (\emph{3.170})}&\specialcell{3.203\\ (\emph{3.274})}&\specialcell{3.384\\ (\emph{3.158})}&\specialcell{3.364\\ (\emph{3.158})}&\specialcell{3.426\\ (\emph{3.160})}&\specialcell{3.393\\ (\emph{3.159})}&3.888\newline (--)&3.161\newline (--) \tabularnewline
&5&\specialcell{5.485\\ (\emph{5.781})}&\specialcell{5.293\\ (\emph{5.407})}&\specialcell{5.307\\ (\emph{5.271})}&\specialcell{5.361\\ (\emph{5.241})}&\specialcell{5.289\\ (\emph{5.412})}&\specialcell{5.571\\ (\emph{5.223})}&\specialcell{5.540\\ (\emph{5.224})}&\specialcell{5.642\\ (\emph{5.227})}&\specialcell{5.590\\ (\emph{5.225})}&6.372\newline (--)&5.280\newline (--) \tabularnewline
    \bottomrule\bottomrule \addlinespace[1.5ex]
    \end{tabularx}
    \parbox{\textwidth}{\emph{Notes:} $\hat{s}$ denotes the number of selected variables excluding the constant. `False pos.' and `False neg.' denote the number of falsely included and falsely excluded variables, respectively. `Bias' is the $\ell_1$-norm bias defined as $\sum_{j}|\hat\beta_j-\beta_j|$ for $j=1,\ldots,p$. `RMSE' is the root mean squared error (a measure of in-sample fit) and `RMSPE' is the root mean squared prediction error (a measure of out-of-sample prediction performance); see equation~\eqref{eq:rmse_rmspe}. Post-estimation OLS results are shown in parentheses if applicable. \texttt{cvlasso} results are for 5-fold cross-validation. The oracle estimator applies OLS to all predictors in the true model (i.e., variables 1 to $s$). Thus, the false positive and false negative frequency is zero by design for the oracle. The number of replications is 1,000.} 
    \caption{Monte Carlo simulation for an exactly sparse parameter vector with $p=100$ and $\obs=200$.}\label{tab:mc_design1}
\end{table}

For comparison, we also show results of stepwise regression (for $p=100$ only) and the oracle estimator. Stepwise regression starts from the full model and iteratively removes regressors if the $p$-value is above a pre-defined threshold (10\% in our case). Stepwise regression is known to suffer from overfitting and pre-testing bias. However, it still serves as a relevant reference point due to its connection with \emph{ad hoc} model selection using hypothesis testing and the general-to-specific approach. The oracle estimator is OLS applied to the predictors included in the true model. Naturally, the oracle estimator is expected to show the best performance, but is not feasible in practice since the true model is not known. 

We first summarize the main results for the case where $p=100$; see Table~\ref{tab:mc_design1}. 
AIC and stepwise regression exhibit the worst selection performance, with around 18-20 falsely included predictors on average. While AIC and stepwise regression achieve the lowest RMSE (best in-sample fit), the out-of-sample prediction performance is among the worst---a symptom of over-fitting. It is interesting to note that the RMSE of AIC and stepwise regression are lower than the RMSE of the oracle estimator. The corrected AIC improves upon the standard AIC in terms of bias and prediction performance. 

Compared to AIC$_c$, the BIC-type information criteria show similar out-of-sample prediction and better selection performance. While the EBIC performs only marginally better than BIC in terms of false positives and bias, we expect the relative performance of BIC and EBIC to shift in favour of EBIC as $p$ increases relative to $\obs$. 
5-fold CV with the lasso behaves very similarly to the AIC$_c$ across all measures. The rigorous lasso, rigorous square-root lasso and EBIC exhibit overall the lowest false positive rates, whereas rigorous methods yield slightly higher RMSE and RMSPE than IC and CV-based methods. However, post-estimation OLS (shown in parentheses) applied to the rigorous methods improves upon first-step results, indicating that post-estimation OLS successfully addresses the shrinkage bias from rigorous penalization. The performance difference between X-dependent and X-independent penalty choices are minimal overall.

\begin{table}
    \centering\singlespacing
    \newcolumntype{C}{>{\centering\arraybackslash}X}
    \newcommand{\myrotate}[1]{\multirow{5}{*}{\rotatebox{90}{#1}\hspace*{-3pt}}}
    \newcommand{\specialcell}[2][t]{\begin{tabular}[#1]{@{}c@{}}#2\end{tabular}}
    \scriptsize
    \begin{tabularx}{\textwidth}{llCCCCC@{\hspace{1.9\tabcolsep}}CCCCCCCC}
    \toprule \midrule
     &$\sigma$&\multicolumn{4}{l}{\lassotwo}&\multicolumn{1}{l}{\cvlasso}&\multicolumn{4}{l}{\rlasso}&{Oracle} \tabularnewline
     & &{AIC}&{AIC$_c$}&{BIC}&{EBIC$_\xi$}&&\multicolumn{2}{l}{lasso} &\multicolumn{2}{l}{$\sqrt{\textnormal{lasso}}$}&\tabularnewline
      & & &  & & & &  & \texttt{xdep} & &  \texttt{xdep} &\tabularnewline
    \midrule\addlinespace[1.5ex]
    \myrotate{$\hat{s}$}&.5&164.38&26.29&21.58&20.58&27.16&20.15&20.17&20.19&20.22&-- \tabularnewline
&1&178.68&26.03&21.53&20.59&27.05&20.24&20.26&20.21&20.24&-- \tabularnewline
&2&187.55&25.95&31.54&20.14&26.61&19.83&19.87&19.70&19.76&-- \tabularnewline
&3&191.44&24.64&92.26&18.48&25.65&18.14&18.20&17.88&17.98&-- \tabularnewline
&5&195.18&23.37&177.00&15.21&23.02&15.05&15.14&14.57&14.73&-- \tabularnewline
\midrule \myrotate{False pos.}&.5&144.38&6.29&1.58&0.58&7.16&0.15&0.17&0.19&0.22&-- \tabularnewline
&1&158.91&6.03&1.54&0.59&7.06&0.24&0.26&0.21&0.24&-- \tabularnewline
&2&169.13&6.34&12.00&0.57&6.97&0.29&0.31&0.22&0.26&-- \tabularnewline
&3&173.49&6.37&74.26&0.51&7.30&0.22&0.23&0.16&0.20&-- \tabularnewline
&5&177.29&7.90&159.41&0.47&7.40&0.21&0.25&0.15&0.18&-- \tabularnewline
\midrule \myrotate{False neg.}&.5&0.00&0.00&0.00&0.00&0.00&0.00&0.00&0.00&0.00&-- \tabularnewline
&1&0.22&0.00&0.00&0.00&0.00&0.00&0.00&0.00&0.00&-- \tabularnewline
&2&1.58&0.39&0.46&0.43&0.37&0.46&0.45&0.53&0.50&-- \tabularnewline
&3&2.05&1.74&2.00&2.04&1.65&2.07&2.04&2.28&2.21&-- \tabularnewline
&5&2.12&4.53&2.41&5.26&4.38&5.16&5.11&5.58&5.45&-- \tabularnewline
\midrule \myrotate{Bias}&.5&\specialcell{19.205\\ (\emph{30.730})}&\specialcell{2.019\\ (\emph{2.338})}&\specialcell{1.913\\ (\emph{1.975})}&\specialcell{1.904\\ (\emph{1.876})}&\specialcell{2.014\\ (\emph{2.388})}&\specialcell{2.099\\ (\emph{1.821})}&\specialcell{2.070\\ (\emph{1.825})}&\specialcell{2.001\\ (\emph{1.829})}&\specialcell{1.982\\ (\emph{1.832})}&1.793\newline (--) \tabularnewline
&1&\specialcell{53.073\\ (\emph{79.744})}&\specialcell{4.029\\ (\emph{4.663})}&\specialcell{3.827\\ (\emph{3.960})}&\specialcell{3.807\\ (\emph{3.763})}&\specialcell{4.030\\ (\emph{4.775})}&\specialcell{4.007\\ (\emph{3.673})}&\specialcell{3.978\\ (\emph{3.678})}&\specialcell{4.002\\ (\emph{3.665})}&\specialcell{3.965\\ (\emph{3.673})}&3.590\newline (--) \tabularnewline
&2&\specialcell{133.674\\ (\emph{191.444})}&\specialcell{7.980\\ (\emph{9.239})}&\specialcell{14.932\\ (\emph{17.226})}&\specialcell{7.512\\ (\emph{7.470})}&\specialcell{7.901\\ (\emph{9.320})}&\specialcell{7.751\\ (\emph{7.376})}&\specialcell{7.716\\ (\emph{7.374})}&\specialcell{7.827\\ (\emph{7.382})}&\specialcell{7.766\\ (\emph{7.375})}&7.190\newline (--) \tabularnewline
&3&\specialcell{220.488\\ (\emph{318.761})}&\specialcell{11.341\\ (\emph{13.494})}&\specialcell{95.127\\ (\emph{123.928})}&\specialcell{10.370\\ (\emph{10.610})}&\specialcell{11.071\\ (\emph{13.410})}&\specialcell{10.539\\ (\emph{10.519})}&\specialcell{10.508\\ (\emph{10.499})}&\specialcell{10.629\\ (\emph{10.631})}&\specialcell{10.571\\ (\emph{10.587})}&10.813\newline (--) \tabularnewline
&5&\specialcell{409.487\\ (\emph{871.303})}&\specialcell{18.294\\ (\emph{68.788})}&\specialcell{365.813\\ (\emph{771.102})}&\specialcell{14.339\\ (\emph{15.526})}&\specialcell{15.709\\ (\emph{19.627})}&\specialcell{14.299\\ (\emph{15.372})}&\specialcell{14.283\\ (\emph{15.325})}&\specialcell{14.367\\ (\emph{15.635})}&\specialcell{14.331\\ (\emph{15.503})}&18.060\newline (--) \tabularnewline
\midrule \myrotate{RMSE}&.5&\specialcell{0.150\\ (\emph{0.105})}&\specialcell{0.460\\ (\emph{0.441})}&\specialcell{0.480\\ (\emph{0.461})}&\specialcell{0.488\\ (\emph{0.467})}&\specialcell{0.462\\ (\emph{0.439})}&\specialcell{0.553\\ (\emph{0.471})}&\specialcell{0.544\\ (\emph{0.470})}&\specialcell{0.525\\ (\emph{0.470})}&\specialcell{0.519\\ (\emph{0.470})}&0.471\newline (--) \tabularnewline
&1&\specialcell{0.207\\ (\emph{0.134})}&\specialcell{0.927\\ (\emph{0.888})}&\specialcell{0.963\\ (\emph{0.926})}&\specialcell{0.980\\ (\emph{0.939})}&\specialcell{0.929\\ (\emph{0.884})}&\specialcell{1.054\\ (\emph{0.944})}&\specialcell{1.044\\ (\emph{0.944})}&\specialcell{1.055\\ (\emph{0.945})}&\specialcell{1.043\\ (\emph{0.944})}&0.946\newline (--) \tabularnewline
&2&\specialcell{0.277\\ (\emph{0.157})}&\specialcell{1.848\\ (\emph{1.770})}&\specialcell{1.816\\ (\emph{1.739})}&\specialcell{1.963\\ (\emph{1.876})}&\specialcell{1.857\\ (\emph{1.767})}&\specialcell{2.075\\ (\emph{1.885})}&\specialcell{2.061\\ (\emph{1.885})}&\specialcell{2.106\\ (\emph{1.887})}&\specialcell{2.082\\ (\emph{1.886})}&1.890\newline (--) \tabularnewline
&3&\specialcell{0.314\\ (\emph{0.150})}&\specialcell{2.772\\ (\emph{2.656})}&\specialcell{1.770\\ (\emph{1.646})}&\specialcell{2.952\\ (\emph{2.817})}&\specialcell{2.780\\ (\emph{2.643})}&\specialcell{3.103\\ (\emph{2.832})}&\specialcell{3.084\\ (\emph{2.832})}&\specialcell{3.156\\ (\emph{2.835})}&\specialcell{3.122\\ (\emph{2.834})}&2.831\newline (--) \tabularnewline
&5&\specialcell{0.357\\ (\emph{0.122})}&\specialcell{4.592\\ (\emph{4.395})}&\specialcell{0.788\\ (\emph{0.567})}&\specialcell{4.964\\ (\emph{4.714})}&\specialcell{4.641\\ (\emph{4.415})}&\specialcell{5.170\\ (\emph{4.735})}&\specialcell{5.140\\ (\emph{4.733})}&\specialcell{5.260\\ (\emph{4.743})}&\specialcell{5.203\\ (\emph{4.739})}&4.713\newline (--) \tabularnewline
\midrule \myrotate{RMSPE}&.5&\specialcell{0.875\\ (\emph{1.165})}&\specialcell{0.541\\ (\emph{0.559})}&\specialcell{0.544\\ (\emph{0.539})}&\specialcell{0.549\\ (\emph{0.532})}&\specialcell{0.542\\ (\emph{0.561})}&\specialcell{0.614\\ (\emph{0.528})}&\specialcell{0.604\\ (\emph{0.528})}&\specialcell{0.584\\ (\emph{0.528})}&\specialcell{0.578\\ (\emph{0.528})}&0.527\newline (--) \tabularnewline
&1&\specialcell{2.079\\ (\emph{2.773})}&\specialcell{1.083\\ (\emph{1.118})}&\specialcell{1.088\\ (\emph{1.078})}&\specialcell{1.097\\ (\emph{1.064})}&\specialcell{1.084\\ (\emph{1.122})}&\specialcell{1.169\\ (\emph{1.058})}&\specialcell{1.159\\ (\emph{1.058})}&\specialcell{1.170\\ (\emph{1.057})}&\specialcell{1.157\\ (\emph{1.058})}&1.056\newline (--) \tabularnewline
&2&\specialcell{4.765\\ (\emph{6.304})}&\specialcell{2.159\\ (\emph{2.228})}&\specialcell{2.320\\ (\emph{2.352})}&\specialcell{2.190\\ (\emph{2.121})}&\specialcell{2.158\\ (\emph{2.230})}&\specialcell{2.296\\ (\emph{2.114})}&\specialcell{2.282\\ (\emph{2.114})}&\specialcell{2.327\\ (\emph{2.114})}&\specialcell{2.303\\ (\emph{2.114})}&2.109\newline (--) \tabularnewline
&3&\specialcell{7.683\\ (\emph{10.352})}&\specialcell{3.232\\ (\emph{3.344})}&\specialcell{5.010\\ (\emph{5.749})}&\specialcell{3.280\\ (\emph{3.180})}&\specialcell{3.227\\ (\emph{3.349})}&\specialcell{3.414\\ (\emph{3.167})}&\specialcell{3.396\\ (\emph{3.166})}&\specialcell{3.466\\ (\emph{3.169})}&\specialcell{3.432\\ (\emph{3.168})}&3.174\newline (--) \tabularnewline
&5&\specialcell{13.782\\ (\emph{27.099})}&\specialcell{5.369\\ (\emph{6.988})}&\specialcell{12.799\\ (\emph{24.532})}&\specialcell{5.422\\ (\emph{5.249})}&\specialcell{5.315\\ (\emph{5.515})}&\specialcell{5.593\\ (\emph{5.227})}&\specialcell{5.565\\ (\emph{5.227})}&\specialcell{5.678\\ (\emph{5.231})}&\specialcell{5.625\\ (\emph{5.229})}&5.285\newline (--) \tabularnewline
    \bottomrule\bottomrule \addlinespace[1.5ex]
    \end{tabularx}
    \parbox{\textwidth}{Stepwise regression is not reported, as it is infeasible if $p>n$. See also notes in Table~\ref{tab:mc_design1}.}
    \caption{Monte Carlo simulation for an exactly sparse parameter vector with $p=220$ and $\obs=200$.}\label{tab:mc_design1_hd}
\end{table}

We also present simulation results for the high-dimensional setting in Table~\ref{tab:mc_design1_hd}. Specifically, we consider $p=220$ instead of $p=100$, while keeping the estimation sample size constant at $\obs=200$. With on average between 164 and 195 included predictors, it is not surprising that the AIC suffers from overfitting. The RMSPE of the AIC exceeds the RMSE by a factor of 5 or more. In comparison, AIC$_c$ and 5-fold cross-validation perform better as model selectors, with a false positive frequency between 6 and 8 predictors. 

Despite the large number of predictors to choose from, EBIC and rigorous methods perform generally well in recovering the true structure. The false positive frequency is below~1 across all noise levels, and the false negative rate is zero if $\sigma$ is 1 or smaller. While the BIC performs similarly to the EBIC for $\sigma=0.5$ and $\sigma=1$, its performance resembles the poor performance of AIC for larger noise levels. The Monte Carlo results in Table~\ref{tab:mc_design1_hd} highlight that EBIC and rigorous methods are well-suited for the high-dimensional setting where $p>\obs$, while AIC and BIC are not appropriate. 

\begin{table}
    \centering\singlespacing\small
\begin{tabular}{llr@{.}lr@{.}l}
\hline\hline
\emph{Method} & \emph{Call} & \multicolumn{4}{l}{\it Seconds} \\
& & \multicolumn{2}{l}{$p=100$} &  \multicolumn{2}{l}{$p=220$}  \\
\hline 
Rigorous lasso  &\rlasso\ \texttt{y\ x}                     &  0&09 &0&24\\ 
\hspace*{.4cm}with X-dependent penalty & \rlasso\ \texttt{y\ x,\ xdep}  &  5&92  &12&73\\ 
Rigorous square-root lasso & \rlasso\ \texttt{y\ x,\ sqrt}          &  0&39  &0&74\\ 
\hspace*{.4cm}with X-dependent penalty & \rlasso\ \texttt{y\ x,\ sqrt\ xdep}   &  3&34  &7&03\\ 
Cross-validation &\cvlasso\ \texttt{y\ x,\ nfolds(5)\ lopt}       & 23&50 &293&93\\ 
Information criteria&\lassotwo\ \texttt{y\ x}                    &  3&06 &44&06\\ 
Stepwise regression &\texttt{stepwise,\ pr(.1):\ reg\ y\ x}       &  4&65 &\multicolumn{2}{c}{--}\\ 
\hline\hline
\multicolumn{6}{l}{\scriptsize PC specification: Intel Core i5-6500 with 16GB RAM, Windows 7.}
\end{tabular}  
\caption{Run time with $p=100$ and $p=220$.}\label{tab:mc_runtime}
\end{table}

The computational costs of each method are reported in Table~\ref{tab:mc_runtime}.
\rlasso with \emph{X-independent} penalty is the fastest method considered. The run-time of lasso and square-root lasso with $p=100$ is 0.1s and 0.4s, respectively.
The computational cost increased only slightly if $p$ is increased to $p=220$. \rlasso with \emph{X-dependent} penalty simulates the distribution of the maximum value of the score vector. This process increases the computational cost of the rigorous lasso to 5.9s for $p=100$ (12.7s for $p=220$).  With an average run-time of 3.1 seconds, \lassotwo is slightly faster than \rlasso with \emph{X-dependent} penalty if $p=100$, but slower in the high-dimensional set-up. Unsurprisingly, $K$-fold cross-validation is the slowest method as it requires the model to be estimated $K$ times for a range of tuning parameters.

\section{Technical notes}\label{sec:technical}

\subsection{Pathwise coordinate descent algorithms}
\texttt{lassopack} implements the elastic net and square-root lasso using coordinate descent algorithms.  The algorithm---then referred to as ``shooting''---was first proposed by \citet{Fu1998} for the lasso, and by Van der Kooij (2007) for the elastic net. \citet{Belloni2011a} and \citet{Belloni2014c} employ the coordinate descent for the square-root lasso, and have kindly provided Matlab code.

Coordinate descent algorithms repeatedly cycle over predictors $j=1,...,p$ and update single coefficient estimates until convergence.  Suppose the predictors are centered, standardized to have unit variance and the penalty loadings are $\psi_j=1$ for all $j$.  In that case, the update for coefficient $j$ is obtained using univariate regression of the current partial residuals (i.e., excluding the contribution of predictor $j$) against predictor $j$. More precisely, the update for the elastic net is calculated as
\[\tilde\beta_j \leftarrow \frac{\mathcal{S}\left(\sum_{i=1}^nx_{ij}(y_i-\tilde{y}^{(j)}_i),\lambda\alpha \right)}{1+\lambda(1-\alpha)}.\]
where $\tilde\beta_j$ denotes the current coefficient estimate, $\tilde{y}^{(j)}_i=\sum_{\ell\neq j}x_{i\ell}\tilde\beta_\ell$ is the predicted value without the contribution of predictor $j$. Thus, since the predictors are standardized, $\sum_{i}x_{ij}(y_i-\tilde{y}^{(j)}_i)$ is the OLS estimate of regressing predictor $j$ against the partial residual $(y_i-\tilde{y}^{(j)}_i)$. The function $\mathcal{S}(a,b)$, referred to as \emph{soft-tresholding operator}, \[ \mathcal{S}(a,b)=\left\{ 
\begin{array}{ll}
     a-b& \textrm{if}\ a>0\ \textrm{and}\ b<|a| \\
     a+b& \textrm{if}\ a<0\ \textrm{and}\ b<|a| \\
     0 & \textrm{if}\ b>|a| \\ 
\end{array}
\right.
\]
sets some of the coefficients equal to zero. The coordinate descent algorithm is spelled out for the square-root lasso in \citet[Supplementary Material]{Belloni2014c}.\footnote{Alexandre Belloni provides MATLAB code that implements the pathwise coordinate descent for the square-root lasso, which we have used for comparison.}

The algorithm requires an initial beta estimate for which the Ridge estimate is used.  If the coefficient path is obtained for a list of $\lambda$ values, \lassotwo starts from the largest $\lambda$ value and uses previous estimates as initial values (`warm starts'). See \citet{Friedman2007,Friedman2010}, and references therein, for further information.

\subsection{Standardization}\label{sec:technical_standardization}
Since penalized regression methods are not invariant to scale, it is common practice to standardize the regressors $x_{ij}$ such that $\sum_ix_{ij}^2=1$ before computing the estimation results and then to un-standardize the coefficients after estimation. We refer to this approach as \emph{pre-estimation standardization}.
An alternative is to standardize \emph{on the fly} by adapting the penalty loadings.
The results are equivalent in theory. In the case of the lasso, setting $\psi_j=(\sum_ix_{ij}^2)^{1/2}$ yields the same results as dividing the data by $\sum_ix_{ij}^2$ before estimation.
Standardization on-the-fly is the default in \texttt{lassopack} as it tends to be faster. 
Pre-estimation standardization can be employed using the \ttt{prestd} option.  The \ttt{prestd} option can lead to improved numerical precision or more stable results in the case of difficult problems; the cost is (a typically small) computation time required to standardize the data.
The \texttt{\underbar{unitl}oadings} option can be used if the researcher does not want to standardize data.
In case the pre-estimation-standardization and standardization-\emph{on-the-fly} results differ, the user can compare the values of the penalized minimized objective function saved in \texttt{e(pmse)} (the penalized MSE, for the elastic net) or \texttt{e(prmse)} (the penalized root MSE, for the sqrt-lasso).

\subsection{Zero-penalization and partialling out}
In many applications, theory suggests that specific predictors have an effect on the outcome variable. Hence, it might be desirable to always include these predictors in the model in order to improve finite sample performance. Typical examples are the intercept, a time trend or any other predictor for which the researcher has prior knowledge. \texttt{lassopack} offers two approaches for such situations:
\begin{itemize}[nosep]
    \item \emph{Zero-penalization:} The \myoption{\underbar{notp}en}{\varlist} option of \texttt{lasso2} and \texttt{cvlasso} allow one to set the penalty for specific predictors to zero, i.e., $\psi_\ell=0$ for some $\ell\in\{1,\ldots,p\}$. Those variables are not subject to penalization and will always be included in the model. \texttt{rlasso} supports zero-penalization through the \myoption{\underbar{pnotp}en}{\varlist} option which accommodates zero-penalization in the rigorous lasso penalty loadings; see below.
    \item \emph{Partialling out:} We can also apply the penalized regression method to the data after the effect of certain regressors has been partialled out. Partialling out is supported by \ttt{lasso2}, \ttt{cvlasso} and \ttt{rlasso} using \myoption{\underbar{par}tial}{\varlist} option. The penalized regression does not yield estimates of the partialled out coefficients directly. Instead, \ttt{lassopack} recovers the partialled-out coefficients by post-estimation OLS.
\end{itemize}

It turns out that the two methods---zero-penalization and partialling out---are numerically equivalent. Formally, suppose we do not want to subject predictors $\ell$ with $\bar{p}>\ell\geq p$ to penalization. The zero-penalization and partialled-out lasso estimates are defined respectively as 
\begin{align}
\bm{\hat\beta}(\lambda)&=  \arg\min  \frac{1}{n} \sum_{i=1}^n\left( y_i - \sum_{j=1}^{\bar{p}}x_{ij}\beta_j - \sum_{\ell=\bar{p}+1}^{p}x_{i\ell}\beta_{\ell}\right)^2 
    + \frac{\lambda}{n} \sum_{j=1}^{\bar{p}}\psi_j|\beta_j| \\ 
\textnormal{and} \qquad\bm{\tilde\beta}(\lambda)&=  \arg\min  \frac{1}{n} \sum_{i=1}^n\left( \tilde{y}_i - \sum_{j=1}^{\bar{p}}\tilde{x}_{ij}\beta_j \right)^2 
    + \frac{\lambda}{n} \sum_{j=1}^{\bar{p}}\psi_j|\beta_j| 
\end{align}
where $\tilde{y}_i=y_i-\sum_{\ell=\bar{p}+1}^{p}x_{i\ell}\hat\delta_{y,\ell}$ and $\tilde{x}_{ij}=x_{ij}-\sum_{\ell=\bar{p}+1}^{p}x_{i\ell}\hat\delta_{j,\ell}$ are the residuals of regressing $y$ and the penalized regressors against the set of unpenalized regressors. The equivalence states that $\hat\beta_j=\tilde\beta_j$ for all $j=1,\ldots,\bar{p}$. The result is spelled out in \citet{Yamada2017} for the lasso and ridge, but holds for the elastic net more generally. 

Either the \texttt{\underbar{par}tial(\varlist)} option or the \texttt{\underbar{notp}en(\varlist)} option can be used for variables that should not be  penalized by the lasso.  The options are equivalent in theory (see above), but numerical results can differ in
practice because of the different calculation methods used.  Partialling-out variables can lead to improved numerical precision or more stable results in the case of difficult problems compared to zero-penalization, but may be slower in terms of computation time.

The estimation of penalty loadings in the rigorous lasso introduces an additional complication that necessitates the \rlasso-specific option \texttt{\underbar{pnotp}pen(\varlist)}.
The theory for the \rlasso penalty loadings is based on the penalized regressors after partialling out the unpenalized variables.
The \texttt{\underbar{pnotp}en(\varlist)} guarantees that the penalty loadings for the penalized regressors are the same as if the unpenalized regressors had instead first been partialled-out.

The \texttt{fe} fixed-effects option is equivalent to (but computationally faster and more accurate than) specifying unpenalized panel-specific dummies.
The fixed-effects (`within') transformation also removes the constant as well as the fixed effects.  The panel variable used by the \texttt{fe} option is the panel variable set by \texttt{xtset}.
If installed, the within transformation uses the fast \textbf{ftools} package by \citet{Correia2016}.

The \texttt{prestd} option, as well as the \texttt{\underbar{notp}en(\varlist)} and \texttt{\underbar{pnotp}en(\varlist)} options, can be used as simple checks for numerical stability by comparing results that should be equivalent in theory.
The values of the penalized minimized objective function saved in \texttt{e(pmse)} for the elastic net and \texttt{e(prmse)} for the square-root lasso may also be used for comparison.

\subsection{Treatment of the constant} 
By default the constant, if present, is not penalized; this is equivalent to mean-centering prior to estimation. The \texttt{\underbar{par}tial(\varlist)} option also partials out the constant (if present). To partial out the constant only, we can specify \texttt{\underbar{par}tial(\_cons)}.  Both \texttt{\underbar{par}tial(\varlist)} and \texttt{fe} mean-center the data; the \texttt{\underbar{nocons}tant} option is redundant in this case and may not be specified with these options.  If the \texttt{\underbar{nocons}tant} option is specified an intercept is not included in the model, but the estimated penalty loadings are still estimated using mean-centered regressors (see the \texttt{center} option).

\section{Acknowledgments}
We thank Alexandre Belloni, who has provided MATLAB code for the square-root lasso, and Sergio Correia for supporting us with the use of \textbf{ftools}. We also thank Christopher F Baum, Jan Ditzen, Martin Spindler, as well as participants of the 2018 London Stata Conference and the 2018 Swiss Stata Users Group meeting for many helpful comments and suggestions. All remaining errors are our own.

\bibliographystyle{sj}
\bibliography{library}

\begin{aboutauthors}
Achim Ahrens is Post-doctoral Research Fellow at The Economic and Social Research Institute in Dublin, Ireland.

Mark E. Schaffer is Professor of Econonomics in the School of Social Sciences at Heriot-Watt University, Edinburgh, UK, and a Research Fellow at the Centre for Economic Policy Research (CEPR), London and the Institute for the Study of Labour (IZA), Bonn.

Christian B. Hansen is the Wallace W. Booth Professor of Econometrics and Statistics at the University of Chicago Booth School of Business.
\end{aboutauthors}

\clearpage

\appendix
\renewcommand{\thesection}{\Alph{section}}

\newpage

\section{Additional Monte Carlo results}\label{sec:moremc}


In this supplementary section, we consider an additional design. Instead of defining $\beta_j$ as either 0 or +1, we let the non-zero coefficients alternate between +1 and -1. That is, we define the sparse parameter vector as $\beta_j = (-1)^j.\mathbbm{1}\{j\leq s\}$ for $j=1,\ldots,p$ with $s=20$. All remaining parameters are as in Section~\ref{sec:mc}, and we consider $p=100$. 

\begin{table}[H]
    \centering\singlespacing
    \newcolumntype{C}{>{\centering\arraybackslash}X}
    \newcommand{\myrotate}[1]{\multirow{5}{*}{\rotatebox{90}{#1}}}
        \newcommand{\specialcell}[2][t]{\begin{tabular}[#1]{@{}c@{}}#2\end{tabular}}
    \scriptsize
    \begin{tabularx}{\textwidth}{llCCCCC@{\hspace{1.9\tabcolsep}}CCCCCCCC}
    \toprule \midrule
     &$\sigma$&\multicolumn{4}{l}{\lassotwo}&\multicolumn{1}{l}{\cvlasso}&\multicolumn{4}{l}{\rlasso}&Step&{Oracle} \tabularnewline
     & &{AIC}&{AIC$_c$}&{BIC}&{EBIC$_\xi$}&&\multicolumn{2}{l}{lasso} &\multicolumn{2}{l}{$\sqrt{\textnormal{lasso}}$}&wise&\tabularnewline
      & & &  & & & &  & \texttt{xdep} &   & \texttt{xdep} &\tabularnewline
    \midrule\addlinespace[1.5ex]
    \myrotate{$\hat{s}$}&.5&77.92&57.07&51.41&4.74&65.90&2.34&2.44&2.03&2.19&37.31&-- \tabularnewline
&1&77.80&51.69&3.90&1.88&60.95&1.65&1.83&1.26&1.51&37.30&-- \tabularnewline
&2&56.31&11.76&1.21&0.28&12.12&0.31&0.41&0.20&0.32&31.68&-- \tabularnewline
&3&26.70&6.40&0.35&0.05&5.28&0.06&0.09&0.04&0.06&27.79&-- \tabularnewline
&5&15.06&4.02&0.12&0.01&3.23&0.01&0.03&0.00&0.02&25.03&-- \tabularnewline
\midrule \myrotate{False pos.}&.5&57.92&37.07&31.44&1.58&45.90&0.32&0.35&0.25&0.27&18.31&-- \tabularnewline
&1&57.85&33.25&1.40&0.59&41.90&0.37&0.36&0.35&0.32&18.65&-- \tabularnewline
&2&41.91&7.50&0.91&0.87&7.95&0.80&0.74&0.86&0.78&19.76&-- \tabularnewline
&3&20.08&4.43&0.94&0.97&3.85&0.96&0.95&0.97&0.96&19.88&-- \tabularnewline
&5&11.77&3.21&0.99&1.00&2.78&0.99&0.99&1.00&0.99&19.68&-- \tabularnewline
\midrule \myrotate{False neg.}&.5&0.00&0.00&0.03&16.80&0.00&17.98&17.91&18.20&18.07&0.00&-- \tabularnewline
&1&0.05&1.56&17.45&18.46&0.95&18.61&18.47&18.91&18.71&0.35&-- \tabularnewline
&2&5.59&15.71&19.19&19.77&15.76&19.75&19.67&19.84&19.74&7.08&-- \tabularnewline
&3&13.29&17.88&19.79&19.96&18.26&19.96&19.94&19.97&19.96&11.08&-- \tabularnewline
&5&16.45&18.82&19.95&20.00&19.07&19.99&19.99&20.00&19.99&13.65&-- \tabularnewline
\midrule \myrotate{RMSE}&.5&\specialcell{0.373\\ (\emph{0.359})}&\specialcell{0.434\\ (\emph{0.386})}&\specialcell{0.464\\ (\emph{0.399})}&\specialcell{1.108\\ (\emph{1.063})}&\specialcell{0.409\\ (\emph{0.372})}&\specialcell{1.208\\ (\emph{1.108})}&\specialcell{1.199\\ (\emph{1.105})}&\specialcell{1.230\\ (\emph{1.119})}&\specialcell{1.216\\ (\emph{1.112})}&0.402\newline (--)&0.474\newline (--) \tabularnewline
&1&\specialcell{0.743\\ (\emph{0.715})}&\specialcell{0.901\\ (\emph{0.807})}&\specialcell{1.418\\ (\emph{1.375})}&\specialcell{1.466\\ (\emph{1.429})}&\specialcell{0.849\\ (\emph{0.769})}&\specialcell{1.519\\ (\emph{1.429})}&\specialcell{1.510\\ (\emph{1.420})}&\specialcell{1.534\\ (\emph{1.452})}&\specialcell{1.524\\ (\emph{1.436})}&0.801\newline (--)&0.944\newline (--) \tabularnewline
&2&\specialcell{1.646\\ (\emph{1.577})}&\specialcell{2.110\\ (\emph{2.041})}&\specialcell{2.280\\ (\emph{2.255})}&\specialcell{2.316\\ (\emph{2.307})}&\specialcell{2.126\\ (\emph{2.051})}&\specialcell{2.328\\ (\emph{2.302})}&\specialcell{2.326\\ (\emph{2.294})}&\specialcell{2.331\\ (\emph{2.312})}&\specialcell{2.328\\ (\emph{2.300})}&1.629\newline (--)&1.890\newline (--) \tabularnewline
&3&\specialcell{2.770\\ (\emph{2.690})}&\specialcell{3.064\\ (\emph{3.003})}&\specialcell{3.199\\ (\emph{3.188})}&\specialcell{3.214\\ (\emph{3.212})}&\specialcell{3.103\\ (\emph{3.043})}&\specialcell{3.218\\ (\emph{3.210})}&\specialcell{3.217\\ (\emph{3.207})}&\specialcell{3.218\\ (\emph{3.213})}&\specialcell{3.217\\ (\emph{3.210})}&2.457\newline (--)&2.835\newline (--) \tabularnewline
&5&\specialcell{4.731\\ (\emph{4.636})}&\specialcell{4.988\\ (\emph{4.923})}&\specialcell{5.123\\ (\emph{5.117})}&\specialcell{5.131\\ (\emph{5.131})}&\specialcell{5.033\\ (\emph{4.968})}&\specialcell{5.133\\ (\emph{5.130})}&\specialcell{5.132\\ (\emph{5.128})}&\specialcell{5.133\\ (\emph{5.132})}&\specialcell{5.133\\ (\emph{5.129})}&4.099\newline (--)&4.733\newline (--) \tabularnewline
\midrule \myrotate{RMSPE}&.5&\specialcell{0.638\\ (\emph{0.684})}&\specialcell{0.622\\ (\emph{0.633})}&\specialcell{0.638\\ (\emph{0.618})}&\specialcell{1.143\\ (\emph{1.112})}&\specialcell{0.615\\ (\emph{0.654})}&\specialcell{1.226\\ (\emph{1.138})}&\specialcell{1.218\\ (\emph{1.134})}&\specialcell{1.247\\ (\emph{1.152})}&\specialcell{1.234\\ (\emph{1.143})}&0.623\newline (--)&0.527\newline (--) \tabularnewline
&1&\specialcell{1.279\\ (\emph{1.369})}&\specialcell{1.260\\ (\emph{1.291})}&\specialcell{1.468\\ (\emph{1.455})}&\specialcell{1.500\\ (\emph{1.482})}&\specialcell{1.245\\ (\emph{1.312})}&\specialcell{1.542\\ (\emph{1.483})}&\specialcell{1.535\\ (\emph{1.473})}&\specialcell{1.555\\ (\emph{1.505})}&\specialcell{1.546\\ (\emph{1.490})}&1.258\newline (--)&1.057\newline (--) \tabularnewline
&2&\specialcell{2.436\\ (\emph{2.628})}&\specialcell{2.297\\ (\emph{2.374})}&\specialcell{2.318\\ (\emph{2.324})}&\specialcell{2.336\\ (\emph{2.339})}&\specialcell{2.299\\ (\emph{2.372})}&\specialcell{2.342\\ (\emph{2.340})}&\specialcell{2.341\\ (\emph{2.337})}&\specialcell{2.344\\ (\emph{2.342})}&\specialcell{2.342\\ (\emph{2.339})}&2.595\newline (--)&2.107\newline (--) \tabularnewline
&3&\specialcell{3.378\\ (\emph{3.570})}&\specialcell{3.234\\ (\emph{3.322})}&\specialcell{3.237\\ (\emph{3.247})}&\specialcell{3.240\\ (\emph{3.243})}&\specialcell{3.233\\ (\emph{3.303})}&\specialcell{3.241\\ (\emph{3.244})}&\specialcell{3.241\\ (\emph{3.244})}&\specialcell{3.241\\ (\emph{3.243})}&\specialcell{3.241\\ (\emph{3.244})}&3.821\newline (--)&3.163\newline (--) \tabularnewline
&5&\specialcell{5.341\\ (\emph{5.577})}&\specialcell{5.187\\ (\emph{5.306})}&\specialcell{5.162\\ (\emph{5.172})}&\specialcell{5.161\\ (\emph{5.162})}&\specialcell{5.176\\ (\emph{5.273})}&\specialcell{5.161\\ (\emph{5.163})}&\specialcell{5.161\\ (\emph{5.164})}&\specialcell{5.161\\ (\emph{5.161})}&\specialcell{5.161\\ (\emph{5.163})}&6.245\newline (--)&5.285\newline (--) \tabularnewline
    \bottomrule\bottomrule \addlinespace[1.5ex]
    \end{tabularx}
    \parbox{\textwidth}{See notes in Table~\ref{tab:mc_design1}.}
    \caption{Monte Carlo simulation for exactly sparse parameter vector with alternating~$\beta_j$.}\label{tab:mc_design1_alter}
\end{table}

The results are reported in Table~\ref{tab:mc_design1_alter}. Compared to the base specification in Section~\ref{sec:mc}, the model selection performance deteriorates drastically. The false negative rate is high across all methods. When $\sigma$ is equal to 2 or larger, BIC-type information criteria and rigorous methods often select no variables, whereas AIC and stepwise regression tend to overselect. 

On the other hand, out-of-sample prediction can still be satisfactory despite the poor selection performance. For example, at $\sigma=2$, the RMSPE of cross-validation is only 9.0\% above the RMSPE of the oracle estimator (2.3 compared to 2.11), even though only 4.2 predictors are correctly selected on average. The Monte Carlo results highlight an important insight: model selection is generally a difficult task. Yet, satisfactory prediction can be achieved without perfect model selection.

\end{document}